\documentclass[12pt,a4paper]{iopart}
\usepackage{graphicx}
\usepackage{color}
\expandafter\let\csname equation*\endcsname\relax
\expandafter\let\csname endequation*\endcsname\relax
\usepackage{amsmath}
\usepackage{multirow}
\usepackage{epsfig}
\usepackage{epstopdf}
\usepackage{amssymb}
\usepackage{cancel}
\usepackage{float}
\usepackage{nameref}
\usepackage[utf8]{inputenc}
\usepackage{textcomp}
\usepackage{multicol}
\usepackage{indentfirst}
\usepackage{tabularx}
\usepackage{longtable}
\usepackage{setspace}
\usepackage{framed}
\usepackage{subcaption}
\usepackage{soul}
\usepackage{sidecap}
\usepackage{upgreek}
\usepackage{booktabs}
\usepackage[sorting=none, backend=biber, style=numeric-comp, sortcites=true, minbibnames = 10, maxbibnames = 11]{biblatex}

\usepackage[deletedmarkup=xout]{changes} 
\definechangesauthor[color=blue]{SW}
\definechangesauthor[color=orange]{MK}
\definechangesauthor[color=green]{KN}
\definechangesauthor[color=red]{BH}
\definechangesauthor[color=purple]{AD}

\usepackage[breaklinks=true,colorlinks=true,linkcolor=blue,urlcolor=blue,citecolor=blue]{hyperref}

\begin{document}

\title[Nonlocal dynamics of SEs in CCRF discharges]{Nonlocal dynamics of secondary electrons in capacitively coupled radio frequency discharges}

\author{K. Noesges$^{1,\dagger}$, M. Klich$^{2}$, A. Derzsi$^{3}$, B. Horv\'ath$^{3}$, J. Schulze$^{1,4}$, R. P. Brinkmann$^{2}$, T. Mussenbrock$^{1}$, S. Wilczek$^{1}$}

\address{$^1$Chair of Applied Electrodynamics and Plasma Technology, Faculty of Electrical Engineering and Information Technology, Ruhr University Bochum, D-44780, Bochum, Germany}
\address{$^2$Chair for Theoretical Electrical Engineering, Faculty of Electrical Engineering and Information Technology, Ruhr University Bochum, D-44780, Bochum, Germany}
\address{$^3$Institute for Solid State Physics and Optics, Wigner Research Centre for Physics, H-1121 Budapest, Konkoly-Thege Mikl\'os str. 29-33, Hungary}
\address{$^4$Key Laboratory of Materials Modification by Laser, Ion, and Electron Beams (Ministry of Education), School of Physics, Dalian University of Technology, Dalian 116024, People’s Republic of China}
\address{$^\dagger$Corresponding author: katharina.noesges@rub.de}

\date{\today}

\begin{abstract}
In capacitively coupled radio frequency (CCRF) discharges, the interaction of the plasma and the surface boundaries is linked to a variety of highly relevant phenomena for technological processes.
One possible plasma-surface interaction is the generation of secondary electrons (SEs), which significantly influence the discharge when accelerated in the sheath electric field.
However, SEs, in particular electron-induced SEs ($\updelta$-electrons), are frequently neglected in theory and simulations.
Due to the relatively high threshold energy for the effective generation of $\updelta$-electrons at surfaces, their dynamics are closely connected and entangled with the dynamics of the ion-induced SEs ($\upgamma$-electrons).
Thus, a fundamental understanding of the electron dynamics has to be achieved on a nanosecond timescale, and the effects of the different electron groups have to be segregated.
This work utilizes $1d3v$ Particle-in-Cell/Monte Carlo Collisions (PIC/MCC) simulations of a symmetric discharge in the low-pressure regime ($p\,=\, 1\,\rm{Pa}$) with the inclusion of realistic electron-surface interactions for silicon dioxide.
A diagnostic framework is introduced that segregates the electrons into three groups (``bulk-electrons'', ``$\upgamma$-electrons'', and ``$\updelta$-electrons'') in order to analyze and discuss their dynamics.
A variation of the electrode gap size $L_\mathrm{gap}$ is then presented as a control tool to alter the dynamics of the discharge significantly.
It is demonstrated that this control results in two different regimes of low and high plasma density, respectively.
The fundamental electron dynamics of both regimes are explained, which requires a complete analysis starting at global parameters (e.g., densities) down to single electron trajectories.

\end{abstract} 

\pagebreak

\section{Introduction}

Capacitively coupled radio frequency (CCRF) discharges are used in many industrial processes such as etching and sputtering \cite{Lieberman,ChabertBraithwaite,Deposition_Bienholz}.
Especially in the semiconductor industry, low pressures of a few pascal are needed to realize micro- and nanometer-scale electronics.
High voltages of at least several hundreds of volts are required to ensure that highly energetic ions bombard the surface perpendicularly and high-aspect-ratio trenches can be etched.
The gas composition, the neutral gas pressure, the frequency, the voltage and the electrode gap size are parameters which significantly affect the discharge \cite{Striations,IonDynamis,Wilczek_DrivingFrequency,Liu_2022}.
Especially in industry, the reactor design and thus the choice of electrode gap size play an essential role \cite{Hyo_Chang_GapSizeVariation}.
Precise adjustment of the gap size allows controlling process-relevant properties such as the plasma density and the ion flux onto the surface \cite{Fundamental_CCRF,Ion_flux_nonuniformities,IEDF_control}. \par
At low pressures, the electron mean free path can become larger than the electrode gap size ($\lambda_{\rm{m}} \geq  \rm{L}_{\rm{gap}}$).
Electrons propagate through the discharge collisionlessly and impinge on the opposing boundary sheath \cite{Kaganovich_LowPressure,Wilczek_resonance}.
Many studies \cite{Schulze_ElectronBeam, Liu_BRH2011,Liu_BRH2012, Liu_BRH2012_2, Liu_BRH2013, Wilczek_DrivingFrequency, Hyo_Chang_GapSizeVariation} have shown before that the interaction of an electron beam with the opposing sheath is crucial for the electron dynamics, suggesting that the electrode gap size is a possible control parameter of it.
The impingement phase (collapsing or expanding sheath phase) at the opposite electrode of highly energetic electrons accelerated at the expanding sheath phase dictates whether electrons lose or gain energy during their interaction with the opposite boundary sheath.
Due to the low pressure, the impingement phase is directly correlated to the electrode gap size \cite{BastiPhDThesis}.
\par

Another important effect that affects the electron dynamics is the interaction of electrons and heavy particles with boundary surfaces.
These plasma surface interactions are essential in application processes and are known to significantly influence the plasma \cite{EISEE,ChabertBraithwaite, SEE2f, Donko_SEE2f, Korolov_SEE_EAE, Makabe_Petrovic, Daksha_SEE,Braginsky_2012}.
The significance of individual effects on the plasma depends on the surface material and the discharge conditions \cite{Phelps-Gamma}. Investigating plasma surface interactions and their effects on the overall discharge has become increasingly essential and is the basis of several research projects. \cite{PlasmaSurfaceInteraction,PlasmaSurfaceInteractions}.
One possible interaction is the generation of secondary electrons (SEs) at the surfaces.
Ions, electrons, fast neutrals, excited molecules, and photons can cause the emission of SEs from surfaces upon impact \cite{Chapman}.
However, many studies neglect plasma surface interactions and examine the discharge in the $\alpha$-mode.\cite{Belenguer_Boeuf_TransitionRegimes, Schulze_Stochastic, Schulze_ElectronBeam,Turner_Collisionless,Mussenbrock_OhmicStochastic}.
The $\alpha$-mode \cite{Raizer_CCRF} is a characteristic heating mode of CCRF discharges, where electrons gain energy during the expanding sheath phase and ionize the background gas in the bulk region.
When SE emission is considered, often only ion-induced SEs ($\upgamma$-electrons) are included in the discharge models \cite{Daksha_SEE,SEE_Aranka}.
These $\upgamma$-electrons are generated by ions hitting the surface.
At low incident particle energies, the Auger-Meitner effect \cite{Auger_effect,Hagstrum_Theory} lays the theoretical basis of the process.
For high pressures and high driving voltages, it has been shown that the plasma is operated in the $\upgamma$-mode \cite{Belenguer_Boeuf_TransitionRegimes,SEE2f,Donko_SEE2f, Korolov_2f_EAE, Boehm_SEE}.
In the $\upgamma$-mode, the electron dynamics are dominated by the dynamics of the $\upgamma$-electrons inside the boundary sheath regions.
Thus, the important scaling parameter becomes the ratio between the electron mean free path $\lambda_\mathrm{m}$ and the sheath width $s$ rather than the electrode gap $L_\mathrm{gap}$. \par
Industrial applications, however, utilize low pressures and tend to apply high voltages \cite{VWT_Krueger}.
The mean free path of the high-energy $\upgamma$-electrons is too large to allow the system to be driven in the $\upgamma$-mode.
Nevertheless, $\upgamma$-electrons impact the discharge at these conditions.
Due to the constant ion flux towards the surface, $\upgamma$-electrons are released continuously and gain energy from the time dependent sheath potential.
As a result of the large electron mean free path, many $\upgamma$-electrons propagate collisionlessly through the discharge and hit the opposing electrode with several hundreds of eV energy at or around the time of local sheath collapse.
These high-energy electrons bombarding the surface cause a collision cascade inside the material, and electron-induced SEs ($\updelta$-electrons) are emitted from the surface \cite{Sternglass_ModelSEE, Kanaya_ModelSEE,Suszcynsky_ModelSEE, Richterova_ModelSEE}. 
Thereby, it is to be considered that more than one $\updelta$-electron per incident electron can be generated.
In recent studies it has been shown that $\updelta$-electrons are mainly generated during the sheath collapsing phase and at the beginning of the sheath expansion \cite{EISEE,Wang_2021,Wang_2021x}.
The time-dependent generation of a large number of $\updelta$-electrons and the permanent generation of high-energy $\upgamma$-electrons significantly affects the electron dynamics.
Recent studies have shown that in this regime the ratio of mean free path and electrode gap size ($\lambda_\mathrm{m}/L_\mathrm{gap}$) also influences the impingement phase of electron beams \cite{Liu_BRH2011,Liu_BRH2012, Liu_BRH2012_2,Liu_BRH2013,Jiang_BRH}.
Yet, the individual dynamics of the different electron groups, their interplay and the fundamentals of the disparity between the resulting low and high density regimes remain unknown.\par
This work aims to study these dynamics of both ion-induced and electron-induced SEs for a variable electrode gap size.
The investigations are done computationally with the Particle-In-Cell/Monte Carlo collisions (PIC/MCC) approach.
It is a commonly used kinetic and self-consistent approach for low-pressure capacitively coupled plasmas (CCPs) \cite{Birdsall-PIC, Verboncoeur-PIC, PIC-Zoli,Hockney_Eastwood_ComputerSimulations,Birdall_Langdon_PlasmaViaSimulations,donko2021edupic}.
A diagnostic tool is introduced that divides the electrons into groups of bulk-, $\upgamma$- and $\updelta$-electrons to investigate their influence, separately.
Energy-dependent surface coefficients (SCs) \cite{Vaughan_FormularSEE,Vaughan_NewFormularSEE} for $\rm{SiO}_2$ electrodes are included as proposed by Horv\'ath et al \cite{EISEE,EISEE2018}.
The electron-surface interaction model includes electron-induced SE emission (emission of true SEs, i.e $\delta$-electrons) as well as elastic and inelastic reflection of electrons at the electrode.
The ion-induced SE emission coefficient ($\upgamma$-coefficient) is assumed to be constant.
Both the $\upgamma$-coefficient and the electron reflection coefficient ($r$) are varied to study their influence on the discharge.
Details on the simulation and the SCs are found in section 2.\par
The variation of the electrode gap shows the expected result of a modulated impingement phase of high-energy electrons at the boundary sheath. The influence of the different electron groups on global parameters, such as the ionization rate, is investigated.
It is shown that secondary electrons strongly contribute to the ionization rate, which individual distribution depends on the electrode gap size (for details, see sec. 3.1 - 3.4). At smaller electrode gap sizes, many high-energy $\updelta$-electrons are found in the
system contributing to the ionization process. In contrast, at large electrode gap
sizes, the $\upgamma$-electrons become more important for the ionization process due to their very high power
consumption. Furthermore, the total electron density increases by a factor of six when the $\rm{SiO}_2$ model is used.
A spatially and temporally resolved analysis of the individual high-energy electron densities reveals a "banana-shaped" beam structure formed by secondaries.
This structure is an accumulation of secondary electrons with different velocities, passing each other throughout the discharge.
This phenomenon is analog to caustics in optics \cite{causticoptics}.
The electrode gap size works as a control parameter for the impingement phase of this banana-shaped structure.
It is demonstrated that precise control of the interaction of these fast electrons with the boundary sheath is essential to the electron confinement.
Furthermore, it is shown that the interplay of the SE groups leads to the counter-intuitive finding that well-confined high-energy SEs are found at low plasma densities (cf. small gap sizes). 
Nevertheless, at high plasma densities (cf. large gap sizes), more SEs are found in the system in total, which significantly increases the electron density.
These findings are supported by a detailed analysis of the electron power gain and loss dynamics, where the disparities between bulk electrons and secondaries are emphasized.
Yet, the full dynamics of the secondaries is solely unveiled when the underlying superposed single particle trajectories are discussed (details in sec. 3.5 - 3.6).
The paper is concluded and summarized in section 4.\par

\section{Simulation method and discharge conditions}
The electron dynamics of a symmetric CCRF discharge are studied using the benchmarked $1d3v$ PIC/MCC simulation code called \textit{yapic1D} \cite{PIC-Benchmark}. It is a kinetic electrostatic plasma simulation in which particles are traced within an equidistant Cartesian grid with a cell size of $\Delta x = L_{\rm{gap}}/N_{\rm{cells}}$. The number of cells $N_{\rm{cells}}$ is adjusted so that the Debye length $\lambda_{D}$ \cite{PIC-Benchmark,Birdsall-PIC,donko2021edupic} is resolved.
The plane and parallel electrodes of the CCP are treated as infinite \cite{Wilczek_Tutorial}.
A harmonic waveform of $V(t)\,=\,V_{\rm{0}}\, \sin{(2\pi ft})$ with $V_{\rm{0}}\,=\,500\,\rm{V}$ and $f\,=\,27.12\,\rm{MHz}$ is applied to the electrode at $x\,=\,0\,\rm{mm}$, while the opposite electrode is grounded. The time step is calculated as $\Delta t\,=\,(fN_{\rm{tspc}})\textsuperscript{-1}$, with $N_{\rm{tspc}}$ being the number of time steps per cycle. $N_{\rm{tspc}}$ is chosen to satisfy the requirements regarding the electron plasma frequency \cite{PIC-Benchmark,Birdsall-PIC,Wilczek_Tutorial,donko2021edupic} and to 
ensure that the Courant-Friedrichs-Lewy condition \cite{Courant} is always fulfilled. The concept of superparticles, each representing a large number of particles, is used to lower the computational load of the system \cite{Hockney_Eastwood_ComputerSimulations,Birdall_Langdon_PlasmaViaSimulations, Birdsall-PIC, PIC_Superparticles,donko2021edupic}. The following simulation results were computed with $10^{5}$ superparticles for ions and electrons.
In these simulations, collisions are treated with the MCC \cite{Birdsall-PIC,PIC-Benchmark,donko2021edupic} method combined with a null collision scheme \cite{PIC-Benchmark,SkullerudNull-Collision,Null-Collision}.
Argon is used as a background gas and modeled with the cross section set of Phelps \cite{XsextionPhelps1,XsextionPhelps2,Xsextion1,Xsextion2}. Electron-neutral collisions (elastic scattering, excitation and ionization) as well as ion-neutral collisions (isotropic and backward scattering) are considered \cite{lxcat}.
The electrode gap size $L_{\rm{gap}}$ is varied between $L_\mathrm{gap} = 25\, \text{--}\, 60\,\mathrm{mm}$, while the gas pressure of $p\,=\,1\,\rm{Pa}$ and the gas temperature of $T_{\rm{g}}\,=\, 400\,\rm{K}$ are kept constant. The effect of gas heating is neglected.


\begin{table}[htbp]
    \footnotesize
    \centering
    \begin{tabular}{p{0.25\textwidth}|p{0.15\textwidth} | p{0.15\textwidth} | p{0.1\textwidth}}
        \textbf{Surface models (SMs)}     &$r$                 &$\upgamma$  & $\updelta$	\\
        \specialrule{2pt}{1pt}{1pt}
        SM A	                    & 0.0              & 0.0      & 0.0     \\
        SM B	                    & 0.2              & 0.2      & 0.0     \\
        SM C	                    & $r(\epsilon)$    & 0.2      & $\updelta(\epsilon)$ \\
        SM D	                    & $r(\epsilon)$    & 0.3      & $\updelta(\epsilon)$ \\
        SM E	                    & $r(\epsilon)$    & 0.4      & $\updelta(\epsilon)$ \\
    \end{tabular}
    \caption{Overview of the different surface models. $\qquad \qquad \qquad \qquad \qquad \qquad \qquad$}
    \label{tab:surfacemodel}	
\end{table}

Table \ref{tab:surfacemodel} shows an overview of the surface models used in this work. Electrons hitting the surface can be reflected elastically or inelastically. The reflection model is varied in the following way: A constant reflection coefficient $r$ represents only elastic reflection. Its value is either 0 (no reflection), or 0.2 \cite{Kollath}. $\rm{r}(\epsilon)$ represents elastic and inelastic reflection, both calculated by the energy-dependent $\rm{SiO}_2$ model introduced in \cite{EISEE}. The $\upgamma$-coefficient is varied between $0.0\,-\,0.4$, where the value $\upgamma = 0.4$ corresponds to a $\rm{SiO}_2$ surface \cite{Gamma_SiO2}. Electrons that impinge on the surface can also generate SEs. The probability of generating $\updelta$-electrons is given by the energy-dependent electron-induced secondary electron emission coefficient $\updelta (\epsilon)$, which can be larger than $1.0$ for energies above $\approx\,50\,$eV. 
Furthermore, a diagnostic function is introduced to the simulation, separating the electrons into three groups: bulk-electrons, $\upgamma$-electrons and $\updelta$-electrons.
An electron is considered a bulk-electron if it is generated via ionization by another electron.
The $\upgamma$-electrons and $\updelta$-electrons are electrons that are created by the respective surface process. The electrons remain in their groups until they get lost at the electrodes.
This diagnostic aims to trace different groups of electrons and gain insight into their individual dynamics and their contribution to the ionization rate.

\section{Results}
\label{section:Results}
\subsection{Electrode gap variation}
\label{subsection:GapSizeVariation}

\begin{figure}[h!]
    \centering
    \includegraphics[width =0.7\textwidth]{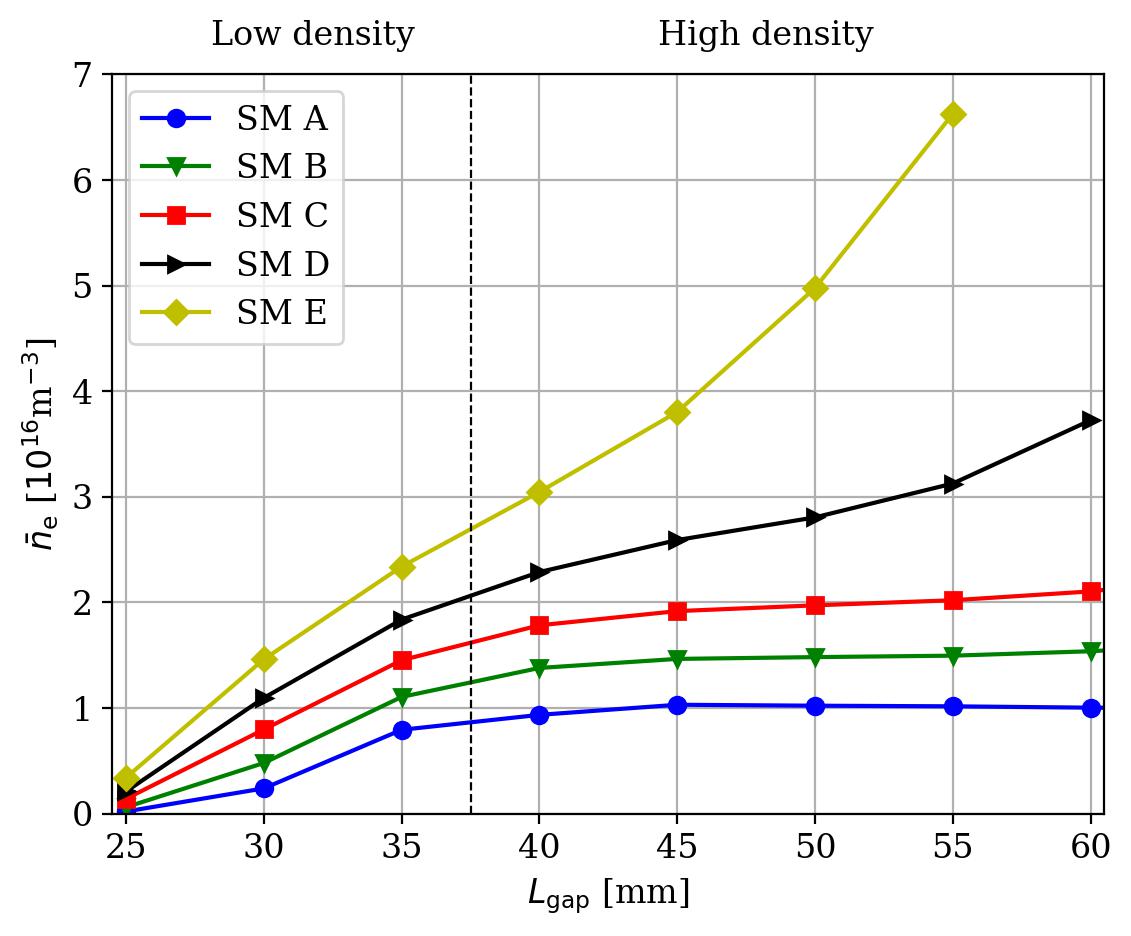}
    \caption{Space- and time-averaged electron density $\bar{n}_\mathrm{e}$ as a function of the gap size $L_\mathrm{gap}$ and the SMs (cf., tab. \ref{tab:surfacemodel}) obtained from the PIC/MCC simulations. The vertical black dashed line represents the subdivision into the low and high electron density regimes. The discharge conditions are: $p\,=\,1.0 \  \rm{Pa}\ [\rm{Ar}],$ $V_0\,=\,500 \  \rm{V},$ $  f \,=\,27.12\ \rm{MHz}$.}
    \label{fig:gapsizevariationSurfaceCoef}	
\end{figure}

In this section, the effects of various electrode gap sizes on the plasma are studied for different surface models to highlight the impact of secondary electrons on the discharge. Figure \ref{fig:gapsizevariationSurfaceCoef} displays the space- and time-averaged electron density as a function of the gap size $L_\mathrm{gap} = 25\, \text{--}\, 60\,\mathrm{mm}$. The different colors represent the results obtained using the previously introduced surface models (SMs, see table \ref{tab:surfacemodel}). 
In SM A (blue), no secondary electrons are considered (i.e., $\upgamma\,=\,0.0, \updelta\,=\,0.0$) and the reflection of electrons at the electrode is neglected ($r\,=\,0.0$). The latter is equivalent to a sticking coefficient of $s=1.0$.
The green line represents SM B, in which the electron reflection coefficient is $r\,=\,0.2$ and solely the ion-induced secondary electrons ($\upgamma$-electrons) are considered in the PIC/MCC simulation, those induced by electrons are neglected ($\upgamma\,=\,0.2, \updelta\,=\,0.0$).
The three remaining lines demonstrate the impact of SM C, D and E (red, black and yellow, respectively). For these models, energy-dependent coefficients for both the reflection of the electrons at the electrode ($r\,=\,r(\epsilon)$) and the emission of the electron-induced secondary electrons ($\updelta$-electrons) ($\updelta\,=\,\updelta(\epsilon)$) are considered \cite{EISEE}.
The models differ in the linearly increased but constant $\upgamma$-coefficient (red: $\upgamma\,=\,0.2$, black: $\upgamma\,=\,0.3$, yellow: $\upgamma\,=\,0.4$).
At the smallest gap size ($L_{\rm{gap}}\,=\,25\,\rm{mm}$), the electron density has the lowest value for all surface models (e.g.,  SM A: $\bar{n}_\mathrm{e} = 3.7\,\cdot\,10^{14} \,\rm{m}^{-3}$, or SM E: $\bar{n}_\mathrm{e} = 3.4\,\cdot\,10^{15} \,\rm{m}^{-3}$). Increasing the electrode gap size to $L_{\rm{gap}}\,=\,35\,\rm{mm}$, the averaged electron density grows by approximately one order of magnitude (e.g., SM A: $\bar{n}_\mathrm{e} = 7.9\,\cdot\,10^{15} \,\rm{m}^{-3}$, or SM E: $\bar{n}_\mathrm{e} = 2.4\,\cdot\,10^{16} \,\rm{m}^{-3}$). The subsequent trend for a further increase of the electrode gap differs between SMs A, B, and C (blue, green, red) and SMs D and E (black, yellow).
For the former SMs, the averaged electron density has an approximately squareroot-shaped trend, and forms a plateau for higher gap sizes by asymptotically striving towards a certain density (e.g., SM A: $\bar{n}_\mathrm{e} = 1 \times 10^{16}\, \mathrm{m}^{-3}$).
In contrast, the latter SMs exhibit a strictly monotonic and strongly increasing trend.
These different trends are based on the distinct electron dynamics caused by the conditions of the SMs.
Furthermore, the small and the larger electrode gap sizes lead to different regimes of the electron dynamics.
This work's discussion unveils both the influence of the surface models on the electron dynamics as well as its different regimes resulting from the gap size $L_\mathrm{gap}$.

\begin{figure}[h!]
\centering
\includegraphics[width =\textwidth]{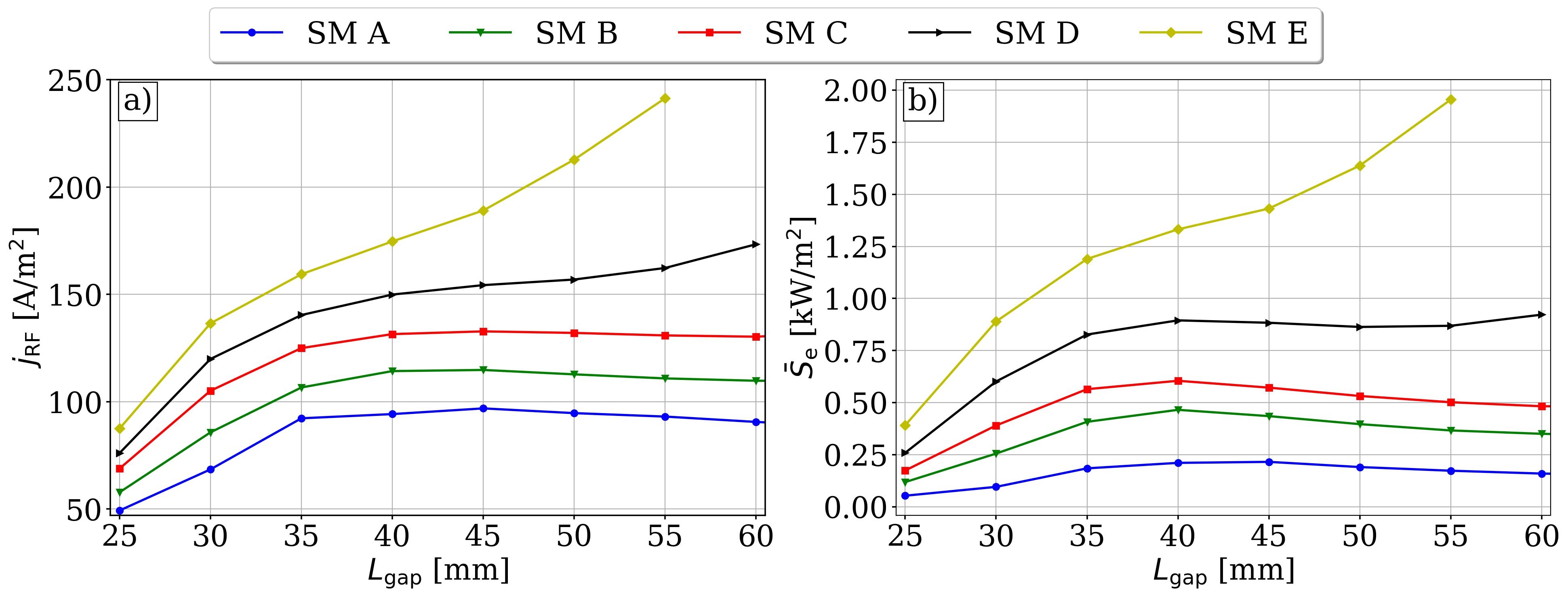}
\caption{First harmonic of the total RF current density $j_{\rm{RF}}$ at the driven electrode obtained from a Fourier analysis (a) and space- and time-averaged electron energy flux $\bar{S}_e$ (b) as a function of the gap size $L_\mathrm{gap}$ and the SMs (cf., tab. \ref{tab:surfacemodel}) obtained from the PIC/MCC simulations. Discharge conditions: $p\,=\,1.0 \  \rm{Pa}\ [\rm{Ar}],$ $V_0\,=\,500 \  \rm{V},$ $f \,=\,27.12\ \rm{MHz}$.}
\label{fig:gapsizevariationSurfaceCoef1}	
\end{figure}

Figure \ref{fig:gapsizevariationSurfaceCoef1}(a) shows the first harmonic of the total RF current density $j_\mathrm{RF}$. This applies for every occurrence of $j_\mathrm{RF}$ at the driven electrode extracted from a Fourier analysis \cite{Wilczek_Current}, where an amplitude phase notation is used.
The current density is plotted as a function of the gap size. As before, the different SMs are represented by different colors and all data are retrieved from PIC/MCC simulations.
Please note that higher harmonics in the RF current hardly contribute to the total current.
Even harmonics cancel each other since it is a symmetrical discharge driven by a sinusoidal voltage \cite{Lieberman,Wilczek_resonance}.
Odd harmonics only range at a very low value (1-2\% of the first harmonic). In general, this figure confirms the trend of the electron density from figure \ref{fig:gapsizevariationSurfaceCoef}. Increasing the SE emission coefficients ($\upgamma$ and $\updelta$) while the RF voltage is kept constant leads to a higher discharge current and, consequently, a higher plasma density.

Figure \ref{fig:gapsizevariationSurfaceCoef1}(b) shows the space- and time-averaged electron energy flux density 
as a function of the gap size and the SCs obtained from the PIC/MCC simulation:
\begin{align}
\bar{S}_{\rm{e}}\,=\,<j_e(x,t)\,\cdot E(x,t)>\,L_{gap}. \label{energyflux}
\end{align}
The electron energy flux density is calculated by the spatially and temporally averaged product of the electron current density and electric field multiplied by the gap size.
The energy flux of the SM A, B, and C increases until $L_{\rm{gap}}\,=\,40\,\rm{mm}$ and then decreases slightly. For $L_{\rm{gap}}\,=\,40\,\rm{mm}$ an optimum energy gain by the electrons is found. The generation of such a plateau has already been observed by   
Liu and co-workers \cite{Liu_BRH2011,Liu_BRH2012, Liu_BRH2012_2,Liu_BRH2013,Jiang_BRH}.
However, no plateau in the electron density as well as in the current density is reached when SM D or E are applied. The reason for this is the large number of SEs in the system.
Horv\'{a}th \textit{et al.} \cite{EISEE} demonstrated that $\updelta$-electrons are mostly generated by the highly energetic $\upgamma$-electrons under such conditions.
This finding applies for the discharges studied here, too.
Due to the increased $\upgamma$-coefficient of SMs D and E, more $\upgamma$- as well as $\updelta$-electrons are generated.
As a direct consequence, the current density keeps increasing resulting in an increasing electron energy flux density for gap sizes above $L_{\rm{gap}}\,=\,50\,\rm{mm}$ for SM D and above $L_{\rm{gap}}\,=\,40\,\rm{mm}$ for SM E.


\subsection{Impingement phase}
\label{subsection:ImpingementPhase}

Some consequences of such an electrode gap size variation on the electron dynamics has already been observed computationally (in PIC/MCC simulations) as well as experimentally by probe measurements and optical emission spectroscopy \cite{Liu_BRH2011,Liu_BRH2012, Liu_BRH2012_2,Liu_BRH2013,Jiang_BRH}.
Utilizing the spatio-temporal dynamics of the ionization rate as a diagnostic, it has been shown that energetic electron beams are accelerated during sheath expansion and reach the opposing sheath at different temporal phases for different electrode gap sizes \cite{Liu_BRH2012_2}.

\begin{figure}[h!]
\centering
\includegraphics[width =\textwidth]{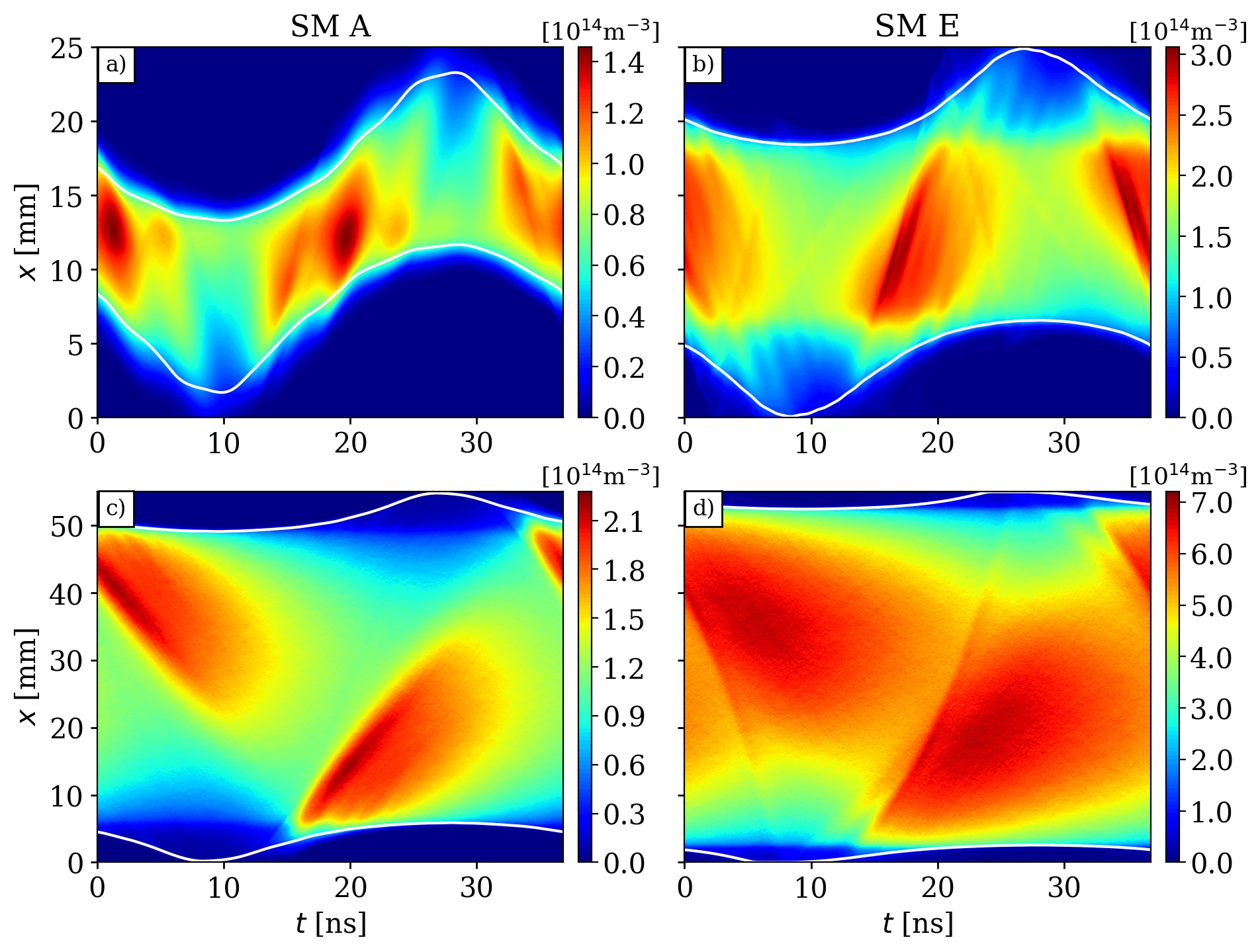}
\caption{Spatio-temporal distribution of the density of electrons with energies above $15.8\,\rm{eV}$.
First row: $L_{\rm{gap}}\,=\,25\,\rm{mm}$, second row: $L_{\rm{gap}}\,=\,55\,\rm{mm}$.
First column: SM A ($r\,=\,0.0,\, \gamma\,=\,0.0,\, \delta\,=\,0.0$), second column: SM E ($r\,=\,r(\epsilon),\, \gamma\,=\,0.4,\, \delta\,=\,\delta(\epsilon)$).
Discharge conditions: $p\,=\,1.0 \  \rm{Pa}\ [\rm{Ar}],$  $V_0\,=\,500 \  \rm{V},$ $f \,=\,27.12\ \rm{MHz}$.}
\label{fig:ElectronDynamics}	
\end{figure}

In this work we observe similar dynamics. The impingement phase of the electron beam at the opposite electrode leads to the different density regimes observed in figure \ref{fig:gapsizevariationSurfaceCoef}.
Furthermore, the dynamics of the electrons are investigated in cases with zero SCs, i.e., neglecting the surface processes and with realistic surfaces.
Figure \ref{fig:ElectronDynamics} shows the spatio-temporal density of electrons with energies above $15.8 \, \rm{eV}$, which is the threshold energy of argon ionization.
In other words, these fast electrons are responsible for sustaining the discharge via ionization. Hence, their dynamics should be understood in detail in order to control the plasma density.
The white lines in figure \ref{fig:ElectronDynamics} represent the electron sheath edge calculated by a criterion proposed by Brinkmann \cite{BrinkmannKriterium, Brinkmann2015, Klich2022}.
The top row of figure \ref{fig:ElectronDynamics} displays the fast electrons at the smallest electrode gap size ($L_{\rm{gap}}\,=\,25\,\rm{mm}$) for SMs A and E.
These electrons are accelerated during the expanding sheath phase.
Due to the low gas pressure ($p\,=\,1.0 \, \rm{Pa}$), the electron mean free path is much larger ($\lambda_m \, \approx \, 45\rm{mm}$) than the electrode gap. Consequently, many electrons traverse through the plasma bulk collisionlessly and interact with the opposing boundary sheath.
However, the small gap size of $L_{\rm{gap}}\,=\,25\,\rm{mm}$ leads to the situation that most of the energetic electrons reach the opposing sheath during the collapsing sheath phase. This scenario is unsuitable for providing an optimal energy gain to the electrons in two aspects \cite{Wilczek_DrivingFrequency}.
First, the interaction with the collapsing sheath phase always leads to a loss of kinetic energy of incoming electrons.
This is because the sheath potential decreases during the residence time of an electron inside the plasma sheath.
Consequently, the reflected electrons return with less energy into the plasma bulk.
This, in turn, influences the ionization process so that the plasma density decreases. Second, due to the decreasing sheath potential, electrons can more easily overcome the full sheath potential and reach the electrode. In this situation, the electron reflection, as well as the $\updelta$-coefficient, become more important.
This is crucial for SM A since no $\updelta$-electrons are considered, and the electrode's reflection probability is set to zero. Therefore, all electrons that hit the electrode are lost, including their kinetic energy. Consequently, the electron density is low in this case. 

Increasing the gap size leads to a shift of the impingement phase of energetic electrons at the opposing sheath.
The bottom row of figure \ref{fig:ElectronDynamics} displays the dynamics of fast electrons at $L_{\rm{gap}}\,=\,55\,\rm{mm}$ for SMs A and E. 
During the expanding sheath phase, the energetic electrons reach the opposing sheath, where electrons are reflected toward the plasma bulk. This mechanism leads to efficient electron confinement and, therefore, to more ionization and higher electron densities. By comparing the cases with no surface processes (SM A) and the most realistic SCs (SM E), it is found that the density of the fast electrons is strongly increased in the latter case. Especially for $L_{\rm{gap}}\,=\,55\,\rm{mm}$, where the impingement phase is suitable for the electron confinement, the maximum density of the fast electrons is about three times as high (i. e $ {n}_\mathrm{f,e} = 7.0 \,\cdot\,10^{15} \,\rm{m}^{-3}$) as without considering surface processes (i. e $ {n}_\mathrm{f,e} = 2.1 \,\cdot\,10^{15} \,\rm{m}^{-3}$). 
Another important finding in connection with figure \ref{fig:ElectronDynamics} is that a different number of electron beams is present in the subplots of figure \ref{fig:ElectronDynamics}. There are different reasons for the appearance of multiple electron beams.
\begin{itemize}
    \item[i)] Nonlinear Dynamics: figure \ref{fig:ElectronDynamics} (a) and (b) indicate multiple beam structures due to nonlinear dynamics. Because of the low plasma density at $L_{\rm{gap}}\,=\,25\,\rm{mm}$, the local electron plasma frequency is very low. Consequently, electrons cannot react instantaneously to the electric field close to the plasma sheath edge. Thus, multiple electron beams are accelerated during one phase of sheath expansion. These nonlinear dynamics have been well studied in previous papers \cite{Wilczek_resonance, Wilczek_DrivingFrequency, Mussenbrock_OhmicStochastic, Mussenbrock_Nonlinear, Sharma_Wave, Birk_Multibeams, Wilczek_Current, Wilczek_Tutorial} and are therefore not the focus of the present work.

    \item[ii)] Reflected electron beams: figure \ref{fig:ElectronDynamics} (b) demonstrates the reflection of energetic electrons at the opposing sheath during the collapsing phase (e.g., $x \approx 20\, \mathrm{mm}$ and $t \approx 23\, \mathrm{ns}$). This indicates that, although the electrons hit the collapsing phase, a certain fraction of these electrons is confined.

    \item[iii)] Secondary electrons: It is visible in figure \ref{fig:ElectronDynamics}(d) that, in addition to the accelerated electrons during sheath expansion, another slightly curved beam is formed, which hits the opposing fully collapsed sheath.
\end{itemize}
\noindent The latter two of these points above and the influence of secondary electrons with respect to these structures are examined in more detail later.


\begin{figure}[h!]
\centering
\includegraphics[width =\textwidth]{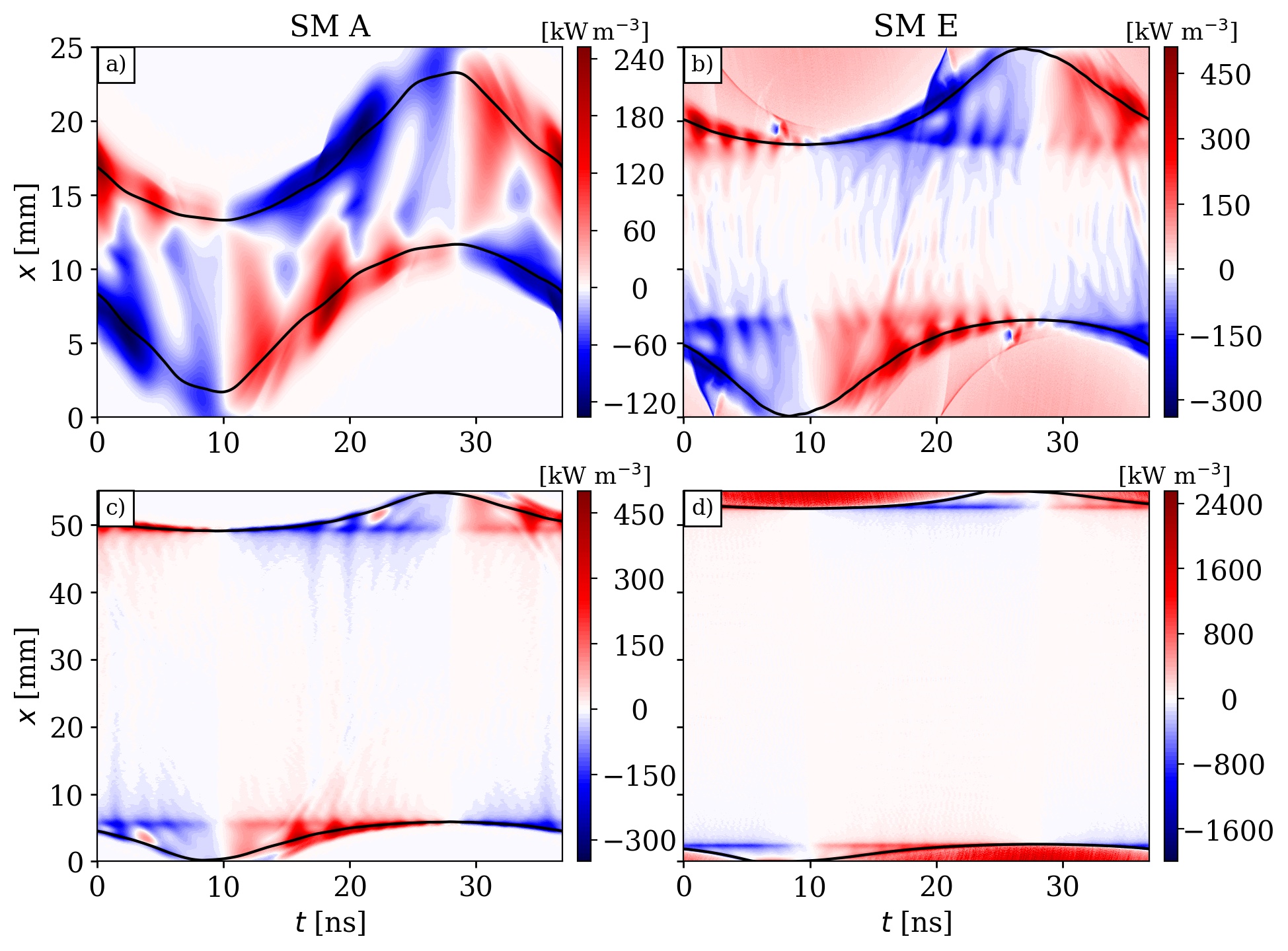}
\caption{Spatio-temporal distribution of the total electron power density.
First row: $L_{\rm{gap}}\,=\,25\,\rm{mm}$, second row: $L_{\rm{gap}}\,=\,55\,\rm{mm}$.
First column: SM A ($r\,=\,0.0,\, \gamma\,=\,0.0,\, \delta\,=\,0.0$), second column: SM E ($r\,=\,r(\epsilon),\, \gamma\,=\,0.4,\, \delta\,=\,\delta(\epsilon)$).
Discharge conditions: $p\,=\,1.0 \  \rm{Pa}\ [\rm{Ar}],$ $V_0\,=\,500 \  \rm{V},$ $f \,=\,27.12\ \rm{MHz}$.
}
\label{fig:ElectronPowerDynamics}	
\end{figure}

Figure \ref{fig:ElectronPowerDynamics} shows the corresponding spatio-temporal distribution of the electron power density:
\begin{align}
\bar{P}_{\rm{e}}\,=\,j_e(x,t)\,\cdot E(x,t). \label{powerdensity}
\end{align}
The power density is calculated as the product of the electron current density times the electric field.
The top row displays the power distribution at the smallest electrode gap size ($L_{\rm{gap}}\,=\,25\,\rm{mm}$) for SM A and E. The bottom row shows it for $L_{\rm{gap}}\,=\,55\,\rm{mm}$. For all four cases, electron power gain (red color) is observed during the expanding sheath phase, where the electrons are accelerated towards the bulk. At the opposing electrode, electron power loss (blue color) can be recognized during the collapsing sheath phase, where electrons are decelerated. Much more electron power gain in the $L_{\rm{gap}}\,=\,55\,\rm{mm}$ cases happens, since the electron beam hits the expanding sheath phase. In figure \ref{fig:ElectronPowerDynamics}, some unique electron power gain and loss structures can be found. There are different reasons for their appearance.

\begin{itemize}
    \item[i)] Nonlinear dynamics: figure \ref{fig:ElectronPowerDynamics} (a) and (b) indicate alternating red and blue or white structures at the expanding sheath edge. These phenomena are linked to the generation of the multiple beams in \ref{fig:ElectronDynamics} (a) and (b) and not the focus of this work \cite{Wilczek_resonance,Wilczek_Current,Mussenbrock_Nonlinear,Birk_Multibeams,OConell_Wave}.

    \item[ii)] Reflected electron beams: In figure \ref{fig:ElectronPowerDynamics} (b) negative ($22\,\rm{mm}\, \leq \, x \, \leq \, 25\,\rm{mm}$, $19\,\rm{ns}\, \leq \, t \, \leq \, 21\,\rm{ns}$ and $x\,\approx\,6\,\rm{mm}$, $t\,\approx\,26\,\rm{ns}$) and positive ($23\,\rm{mm}\, \leq \, x \, \leq \, 25\,\rm{mm}$, $21\,\rm{ns}\, \leq \, t \, \leq \, 23\,\rm{ns}$  and $x\,\approx\,6\,\rm{mm}$, $t\,\approx\,27\,\rm{ns}$) electron power density is found where the reflection of energetic electrons at the opposing sheath occurs\cite{Schulze_ElectronBeam,Schulze_Stochastic,Birk_Multibeams}.

    \item[iii)] Secondary electrons: By comparing SM A and SM E, it is found that additional electron power gain inside the sheath happens. In figure \ref{fig:ElectronPowerDynamics}(d), much more power ($P_{\rm{e,max}}\,\approx\,2500\,\rm{kW\,m^{-3}}$) is coupled into the system than in \ref{fig:ElectronPowerDynamics}(c) ($P_{\rm{e,max}}\,\approx\,500\,\rm{kW\,m^{-3}}$).
    This trend has already been observed in figure \ref{fig:gapsizevariationSurfaceCoef1}(b). With the increased $\gamma$-coefficient and the resulting higher SE density, the absorbed power density grows \cite{EISEE,Intrasheath}. 
\end{itemize}

\noindent The latter two points and the dynamics of the fast electrons with respect to the SEs are investigated in more detail in the following discussion.
To examine the influence of each individual electron group, the electrons are split into groups of bulk-, $\upgamma$- and $\updelta$-electrons. In the following discussion, global parameters are discussed first and then explained based on the individual electron dynamics. Again, the spatio-temporally resolved density of the fast electrons and the electron power density are discussed to demonstrate which group of electrons is responsible for the specific dynamics.

\subsection{Comparison of the surface models}
\label{subsection:ComparisonSMs}
\begin{figure}[h!]
	\centering
	\includegraphics[width =\textwidth,height=17cm]{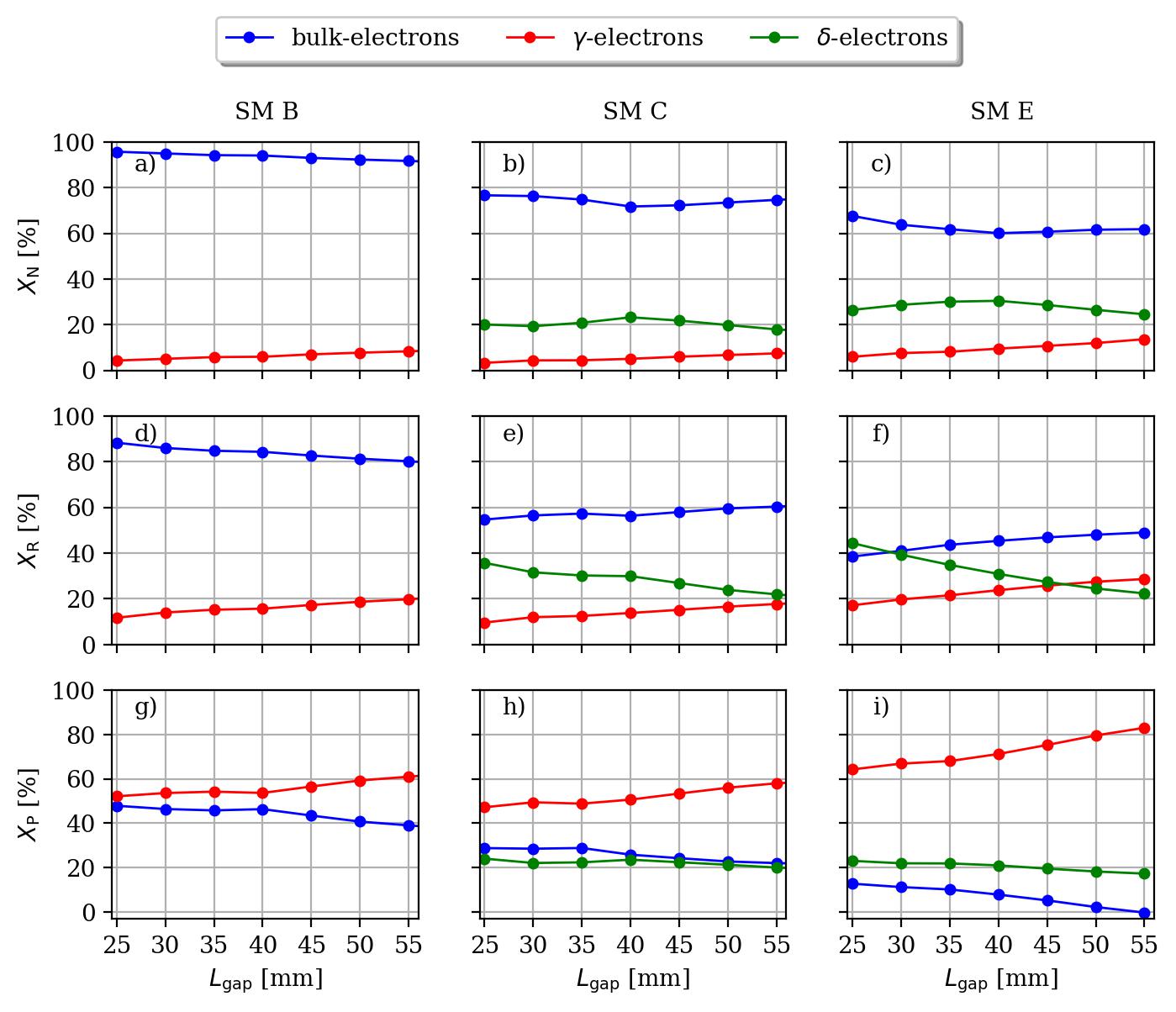}
	\caption{Variation of the gap size $L_{\rm{gap}}\,=\,25\,\rm{mm}\,-\,55\,\rm{mm}$ and the surface models SM B (first column), SM C (second column) and SM E (third column).
     First row: fraction of each electron group density to the total electron density, $X_N$. Second row: contribution of each electron group to the total ionization rate, $X_R$. Third row: contribution of the electron power density of each electron group to the total power dissipated to electrons, $X_P$. Discharge conditions: $p\,=\,1.0 \  \rm{Pa}\ [\rm{Ar}],$ $V_0\,=\,500 \  \rm{V},$ $f \,=\,27.12\ \rm{MHz}$.}
	\label{fig:Matrix}	
\end{figure}

In this section, the influence of specific electron groups on e.g. the plasma dynamics is investigated. For this purpose, the electrons are divided into three groups: bulk-electrons, ion-induced secondary electrons ($\upgamma$-electrons) and electron-induced secondary electrons ($\updelta$-electrons). Based on this classification, characteristic quantities such as electron population, ionization rate, and power coupling are studied as a function of the gap size for different surface models.
It should be noted that SM B (bulk- and $\upgamma$-electrons considered), SM C, and SM E (bulk-, $\upgamma$- and $\updelta$-electrons considered) are compared to study the influence of different considerations of surface coefficients in the simulation on plasma parameters.

For this purpose, three different ratios are considered for each of these groups of electrons: The first ratio $X_{\rm{N}}\, = \, \frac{n_{\rm{e,s}}}{n_{\rm{e}}}$ defines the fraction of each electron group to the total electron density $n_{e}$, where $n_{\rm{e,s}}$ represents the electron density of each electron group. The second ratio $X_{\rm{R}}\, = \, \frac{R_{\rm{e,s}}}{R_{\rm{e}}}$ indicates the contribution of each electron group to the total electron impact ionization rate $R_{\rm{e}}$. $R_{\rm{e,s}}$ is the ionization rate of each electron group.
The third ratio $X_{\rm{P}}\, = \, \frac{P_{\rm{e,s}}}{P_{\rm{e}}}$ provides the share of the electron groups to the total electron power density $P_{\rm{e}}$, where $P_{\rm{e,s}}$ represents the electron power density of each electron group.
Figure \ref{fig:Matrix} shows the previously introduced ratios of the different electron groups for different surface models as a function of the electrode gap size.

Figure \ref{fig:Matrix}(a) displays the population of bulk- and $\upgamma$-electrons for surface model B ($r\,=\,0.2$, $\upgamma\,=\,0.2$ and $\updelta\,=\,0.0$). The fraction of bulk-electrons (blue line) decreases from $96 \%$ to $90 \%$ and the fraction of $\upgamma$-electrons (red line) increases from 4\% to 10\% with increasing electrode gap size. This is related to the variation of the impingement phase and, therefore, to a suitable electron confinement. The individual $\upgamma$-electron dynamics is investigated in more detail later.
Figure \ref{fig:Matrix}(b) shows the trend for SM C, which now includes $\updelta$-electrons (green line). Since the ion-induced secondary electron emission coefficient is still $\upgamma\,=\,0.2$, the trend of the population of $\upgamma$-electrons is the same compared to figure \ref{fig:Matrix}(a). Due to the energy-dependent electron-induced secondary electron emission coefficient $\updelta=\updelta(\epsilon)$, the population of $\updelta$-electrons is in the range of $20 \%$ and bulk- electrons fluctuate in the range between $70 \%$ and $76 \%$. 
In figure \ref{fig:Matrix}(c) (SM E), the $\upgamma$-coefficient is increased to $\upgamma\,=\,0.4$. Consequently, the population of these electrons increases as well. Since $\upgamma$-electrons have high energies as they pass through the boundary plasma sheath, there is in turn a higher probability of generating new $\updelta$-electrons at the opposing electrode. Thus, the population of $\updelta$-electrons increases for SM E up to almost $30 \%$. It should be noted that the variation of the electrode gap size does not drastically affect the population of electrons. However, the variation has a strong impact on the ionization dynamics of the individual electron groups.

Figure \ref{fig:Matrix}(d) shows the relative contribution of bulk- and $\upgamma$-electrons to the total ionization rate for SM B. The bulk-electrons have the largest contribution to the ionization rate ($88 \%$), especially for small gap sizes. The contribution of $\upgamma$-electrons to the ionization rate for $L_{\rm{gap}}\,=\,25\,\rm{mm}$ is approximately $12 \%$ and increases with larger electrode gap sizes.
However, compared to the number of $\upgamma$-electrons, their contribution to the ionization rate is relatively high. Due to their high kinetic energy, some $\upgamma$-electrons can ionize the background gas multiple times while moving through the bulk. 
In figure \ref{fig:Matrix}(e), SM C is analyzed. The share of bulk-electrons to the ionization rate is approximately constant ($\approx 60 \%$) for all gap sizes but lower than in SM B. The influence of the $\upgamma$-electrons over the gap size is similar to SM B due to the constant $\upgamma$-coefficient. In this model, $\updelta$-electrons are also considered, which have a high contribution to the ionization rate ($21\, \text{--}\, 38\,$\%, decreasing with the gap size). 
In figure \ref{fig:Matrix}(f) (SM E), the population and, therefore, the contribution of bulk-electrons to the ionization rate decreases even further. Simultaneously, the secondary electrons become more important for the ionization process. An interplay between $\upgamma$- and $\updelta$-electrons by means of the ionization rate can be found. The contribution of $\upgamma$-electrons for $L_{\rm{gap}}\,=\,25\,\rm{mm}$ is approximately $17 \%$. By increasing the gap size, the contribution of $\upgamma$-electrons rises ($29 \%$). However, especially the $\updelta$-electrons significantly affect the ionization rate and are responsible for the strong increase in the density. For $L_{\rm{gap}}\,=\,25\,\rm{mm}$ the $\updelta$-electrons are essential for the ionization rate in the range of $45 \%$. By increasing the gap size, the contribution of $\updelta$-electrons decreases slowly below the value of the $\upgamma$-electrons ($22 \%$). This transition in the ionization rate will be analyzed in detail in the following.

In figure \ref{fig:Matrix}(g), the share of the bulk- and $\upgamma$-electrons to the total electron power density for SM B is investigated. For the electrode gap size $L_{\rm{gap}}\,=\,25\,\text{--}\,40\,\rm{mm} $ a similar amount of power is coupled into the bulk- and the $\upgamma$-electrons ($\approx 50\, \%$).
More efficient power absorption ($\approx 60\, \%$) by the $\upgamma$-electrons is observed for $L_{\rm{gap}}\,=\,45\,\text{--}\,55\,\rm{mm}$.
This is caused by a suitable impingement phase of the $\upgamma$-electron beam in combination with a larger gap size and, therefore a higher probability of ionization events (cf. following discussion).
By increasing the SCs (fig. \ref{fig:Matrix}(h) and (i)) the power share of bulk-electrons decreases. The power share of the $\upgamma$-electrons in SM C is similar to SM B due to the same $\upgamma$-coefficient. In addition, electron-induced secondary electrons are considered in figure \ref{fig:Matrix}(h). The power absorption of the $\updelta$-electrons is approximately constant ($\approx 20 \%$) for all gap sizes.
In figure \ref{fig:Matrix}(i), the power absorption of the $\updelta$-electrons remains the same. 
In contrast, more and more power is coupled into the $\upgamma$-electrons. Due to the higher $\upgamma$-coefficient, more electrons are released at the electrode and gain energy in the sheath electric field, leading to a higher share in the total electron power flux.  Because the $\upgamma$-electrons are released continuously, they gain energy from the time-dependent sheath potential with a maximum value of $500\,\rm{V}$.
This leads to the large difference between the power density of $\upgamma$-electrons compared to the bulk- and $\updelta$-electrons.
For the case with $L_{\rm{gap}}\,=\,55\,\rm{mm}$, the total electron density is, due to the large number of SEs, very high.
This leads to a smaller sheath width.
The numerous highly energetic secondaries also cause a high electron temperature in the boundary sheath area.
In combination with the reduced thickness of the boundary sheath, a steep temperature gradient is found between the sheath and the bulk region.
This strong temperature gradient lets the power absorption of the bulk-electrons tend to zero \cite{OhmicHeating,PowerAbsorptionNegative}.

In the following discussion, the electron dynamics, with a focus on the SEs, are investigated separately on a nanosecond timescale. The point in time when $\upgamma$- and $\updelta$-electrons are generated is essential because the SEs gain the time-dependent sheath potential.
Below we will understand how the SEs contribute to the ionization process and how they absorb energy compared to bulk-electrons.


\label{section:ElectronDynamics}

\subsection{Electron energy probability distribution functions}
\label{subsection:EEPF}

\begin{figure}[h!]
	\centering
	\includegraphics[width =\textwidth]{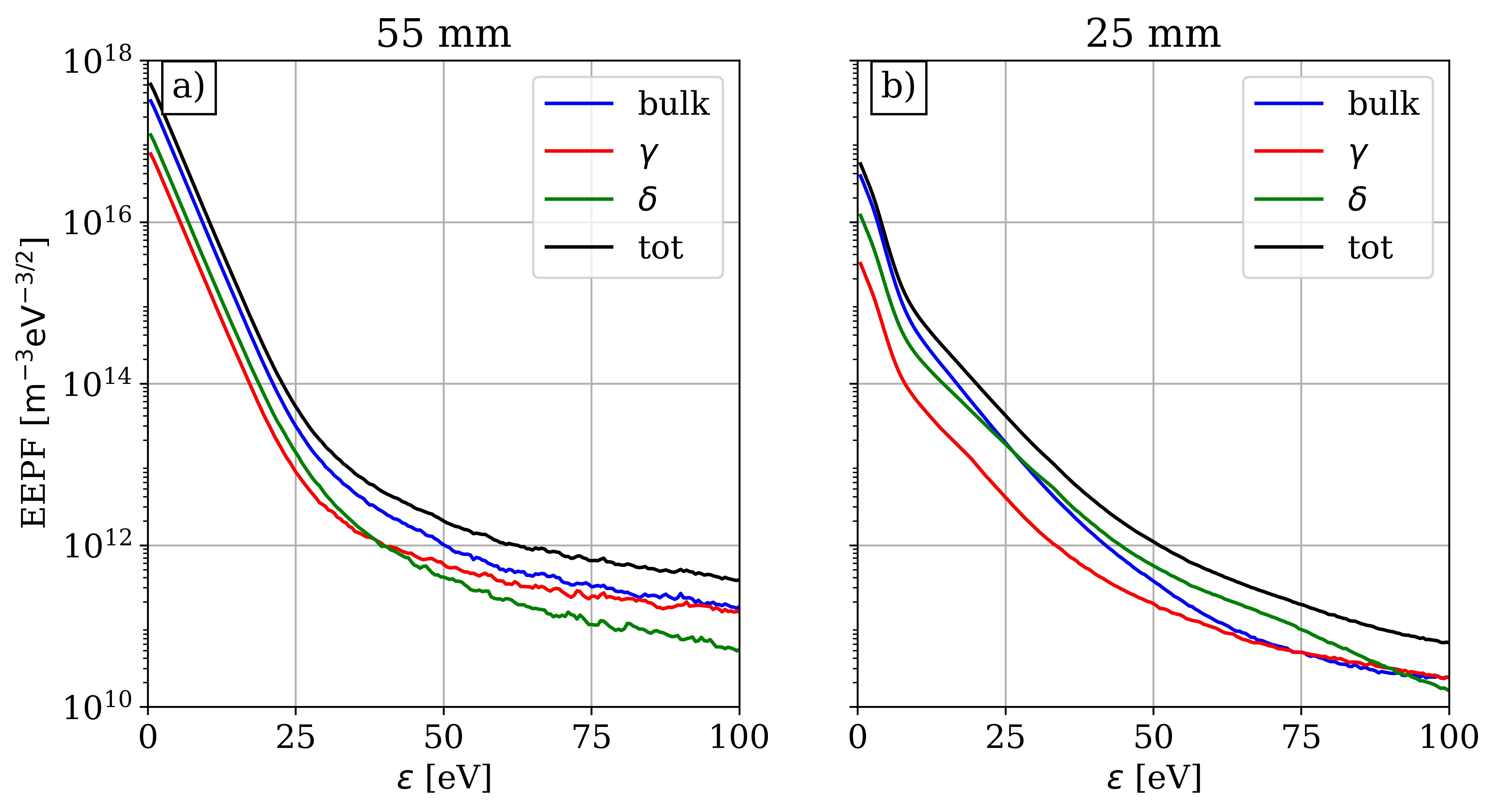}
	\caption{Electron energy probability function (EEPF) in the center of the discharge for each electron group and two gap sizes: $L_{\rm{gap}}\,= \,55\, \rm{mm} \, (a),\  L_{\rm{gap}}\,= \, 25\, \rm{mm} \, (b)$. Discharge conditions: $p\,=\,1.0 \  \rm{Pa}\ [\rm{Ar}],$ $V_0\,=\,500 \  \rm{V},$ $f \,=\,27.12\ \rm{MHz}$, SM E ($r\,=\,r(\epsilon),\, \gamma\,=\,0.4,\, \delta\,=\,\delta(\epsilon)$.}
	\label{fig:EEPF}	
\end{figure}

As pointed out in the previous discussion, the electrode gap size affects the roles of the different electron groups. 
In the following, simulations for both a large electrode gap size ($L_{\rm{gap}}\,=\,55\,\rm{mm}$) and high electron density as well as a small gap size ($L_{\rm{gap}}\,=\,25\,\rm{mm}$) and a low electron density case are demonstratively analyzed.
The discussion will now solely revolve around the simulations applying SM E that yields the most realistic description of the plasma surface interactions. Furthermore, for this surface model, the impact of the secondary electrons is particularly pronounced.

Figure \ref{fig:Matrix}(f) shows an opposing trend for the ionization rates of $\upgamma$- and $\updelta$-electrons as a function of the electrode gap.
Investigating the electron energy probability distribution function (EEPF) reveals the underlying phenomena \cite{ChargedParticlesAndDistributionFunctions}.
Figure \ref{fig:EEPF} shows the time averaged EEPF in the center of the discharge and individually resolved for each electron group (bulk-, $\upgamma$-, and $\updelta$-electrons). The black line corresponds to the total EEPF in the center of the discharge, i.e. the probability function of all types of electrons.
Figure \ref{fig:EEPF}(a) shows the high electron density case ($L_\mathrm{gap} = 55\,$mm) and figure \ref{fig:EEPF}(b) displays the low electron density case ($L_\mathrm{gap} = 25\,$mm). In figure \ref{fig:EEPF}(a), it is most probable to encounter a bulk-electron at any given energy. As the high energy tail of the EEPF is strongly populated by bulk electrons, most of the ionization is caused by bulk electrons in this case (cf., fig. \ref{fig:Matrix}(f)).


The probability distribution of $\updelta$-electrons in the energy interval  $0\,\rm{eV}\, \leq \, \epsilon \, \leq \, 40\,\rm{eV}$ is significantly lower compared to bulk-electrons, and in the range, $40\,\rm{eV}\, \leq \, \epsilon \, \leq \, 100\,\rm{eV}$ their number is the lowest. Thus, the shifted contributions of the respective electron group to the ionization rate (\ref{fig:Matrix}(f)) line up as the high energy parts of their individual EEPFs (in order: bulk- $>$ $\upgamma$- $>$ $\updelta$-electrons). From about $30\,$eV, a high-energy tail can be found in the total EEPF as well as in the distribution function of the individual electron group.
Therefore, it can be concluded that the EEPFs have a bi-Maxwellian-like shape. Reducing the electrode gap size to $L_\mathrm{gap} = 25\,$mm changes the shape of the distribution function. The EEPFs shown in figure \ref{fig:EEPF}(b) appear tri-Maxwellian-like. Three different slopes can be recognized, in the energy range of $0\,\rm{eV}\, \leq \, \epsilon \, \leq \, 10\,\rm{eV}$, in the interval of $10\,\rm{eV}\, \leq \, \epsilon \, \leq \, 60\,\rm{eV}$ and in the range of $60\,\rm{eV}\, \leq \, \epsilon \, \leq \, 100\,\rm{eV}$.
In the energy interval $25\,\rm{eV}\, \leq \, \epsilon \, \leq \, 90\,\rm{eV}$, the $\updelta$-electrons are the most probable electron group, followed by the bulk- and the $\upgamma$-electrons. This order is also reflected in figure \ref{fig:Matrix} (f) and explains their individual contributions to the ionization rate, analogous to the $55\,$mm case.

In the previous discussion only global parameters related to secondary electrons were studied.
It is understood that the ionization dynamics of the electron group depend on the electrode gap size.
In order to understand the full dynamics, it is necessary to consider both spatially and temporally resolved individual dynamics of SEs.


\subsection{Electron dynamics at large electrode gap size}
\label{subsection:LargeGap}

\begin{figure}[h!]
	\centering
	\includegraphics[width =\textwidth]{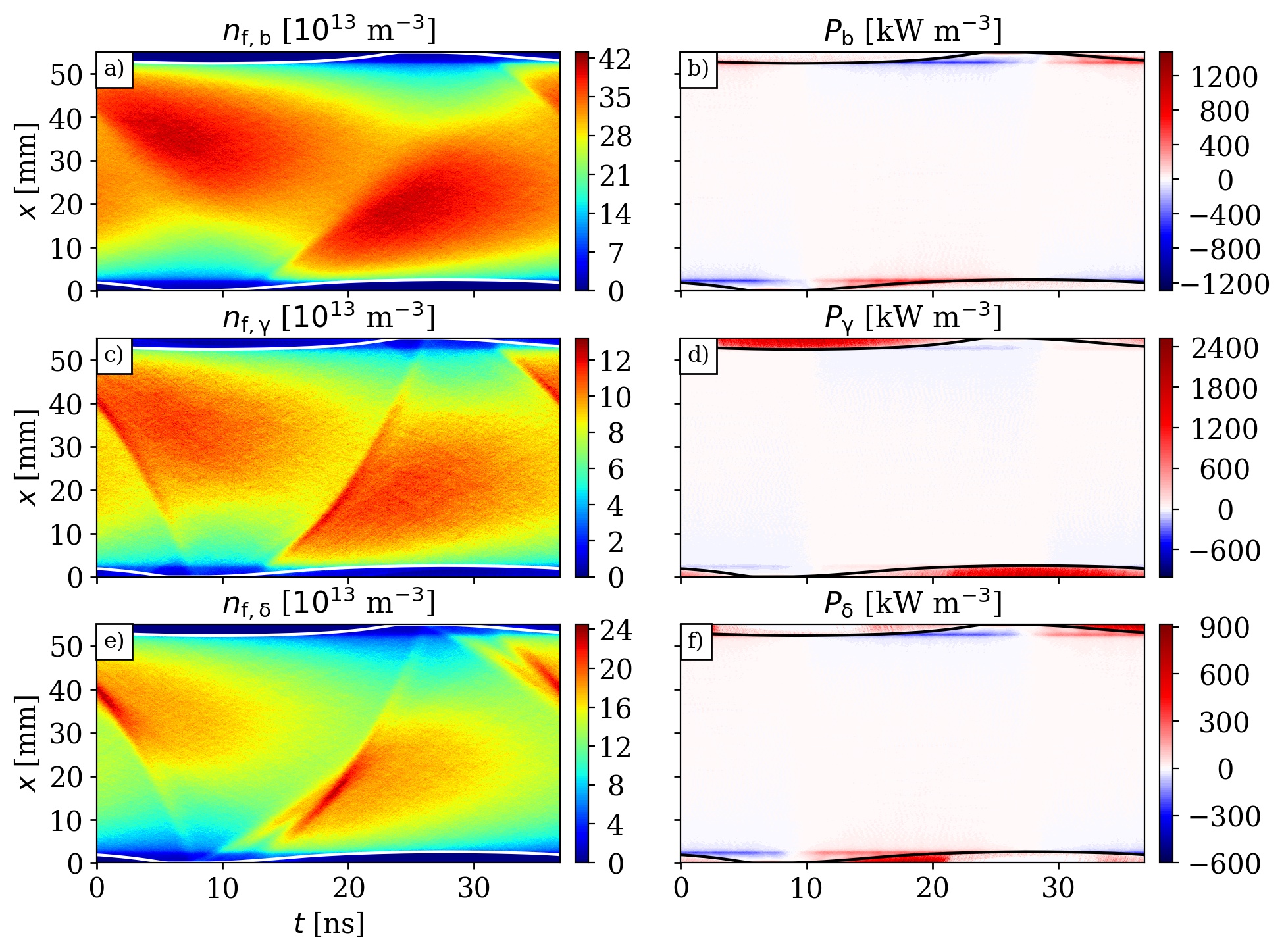}
	\caption{Spatio-temporal distribution of the electron density ($\epsilon\,\geq\,15.8\,\rm{eV}$) and the electron power density of each electron group.
	First column: bulk-electron density (a), $\upgamma$-electron density (c), $\updelta$-electron density (e).
	Second column: bulk-electron power density (b), $\upgamma$-electron power density (d), $\updelta$-electron power density (f). Discharge conditions: $p\,=\,1.0 \  \rm{Pa}\ [\rm{Ar}],$ $V_0\,=\,500 \  \rm{V},$ $f \,=\,27.12\ \rm{MHz},$ $L_{\rm{gap}}\,=\,55\,\rm{mm}$, SM E ($r\,=\,r(\epsilon),\, \gamma\,=\,0.4,\, \delta\,=\,\delta(\epsilon)$).}
	\label{fig:DynamicsE55}	
\end{figure}

First, the dynamics of the three electron groups for $L_{\rm{gap}}\,=\,55\,\rm{mm}$ are studied. The first column of figure \ref{fig:DynamicsE55} shows the spatio-temporal distribution of the individual electron densities with energies above the ionization threshold ($\epsilon\,\geq\,15.8\,\rm{eV}$) within one RF cycle. The second column gives the electron power density for each group. The rows represent the corresponding electron groups, i.e., (a-b) bulk-electrons, (c-d) $\upgamma$-electrons and (e-f) $\updelta$-electrons. Due to symmetry conditions, the same electron dynamics take place at the opposite electrode temporally shifted by half an RF period. In the following the dynamics of electrons are discussed starting from the plasma sheath expansion ($t \geq 10$ ns) at the bottom (driven) electrode ($x=0\,$mm).

Figure \ref{fig:DynamicsE55}(a) shows the density of bulk-electrons. These electrons are accelerated during sheath expansion. The corresponding electron power density in figure \ref{fig:DynamicsE55}(b) indicates that bulk-electrons gain their energy in the sheath electric field during the expansion phase (e.g., $10\,\rm{ns}\, \leq \, t \, \leq \, 30\,\rm{ns}$ at the driven electrode) as well as in the ambipolar electric field in front of the plasma sheath edge \cite{Wilczek_Tutorial,schulze2014effect}. Additionally, bulk-electrons lose their energy hitting the collapsing sheath phase. This trend is frequently observed in experimental works (e.g., phase resolved optical emission spectroscopy) \cite{gans2004prospects,heil2008numerical} as well as in PIC/MCC simulations \cite{Liu_BRH2011,Wilczek_DrivingFrequency}. The density of bulk-electrons with energies above $15.8 \, \rm{eV}$ is much higher than in the case presented in figure \ref{fig:ElectronDynamics}(c) (SM A). The reasoning is that in this scenario secondary electrons contribute strongly to the generation of bulk-electrons.
The resulting deviation is about a factor two. Some of these bulk-electrons can reach the opposing sheath collisionlessly. However, due to the large gap size of $L_{\rm{gap}}\,=\,55\,\rm{mm}$, these electrons reach the opposing sheath during the sheath expansion. Consequently, most of the bulk-electrons do not have enough energy to overcome the sheath potential, reach the electrode, and generate $\updelta$-electrons at the surface. 
In contrast, $\upgamma$-electrons exhibit a different behavior. Figure \ref{fig:DynamicsE55}(d) shows that most of the $\upgamma$-electrons gain their energy inside the sheath. This energy gain mechanism leads  to a maximum energy gain of 500 eV during the fully expanded sheath phase.
Consequently, $\upgamma$-electrons gain most of the power density (cf. fig.  \ref{fig:Matrix} (i)).
Due to collisions with the neutral gas and the resulting isotropisation of the velocity components, the density of fast $\upgamma$-electrons indicates a dynamics similar to the bulk-electrons during sheath expansion (see fig. \ref{fig:DynamicsE55}(c)). This causes a suitable impingement phase for the $\upgamma$-electrons (cf. \ref{fig:Matrix}(g)).
However, in addition to these fundamental dynamics, a banana-shaped structure forms starting at the beginning of sheath expansion ($x \approx 2$ mm, $t \approx 15$ ns). 
This structure reaches the opposing electrode during the fully collapsed sheath phase ($x \approx 25$ mm, $t \approx 25$ ns).
These electrons can now easily overcome the sheath potential
and generate new $\updelta$-electrons, which is illustrated in figure \ref{fig:DynamicsE55}(e).

The density of fast $\updelta$-electrons indicates similar dynamics to the $\upgamma$-electrons. However, shortly after the banana-shaped structure of $\upgamma$- and $\updelta$-electrons reaches the collapsing sheath phase, an additional beam is observed. This beam consists of reflected and newly generated $\updelta$-electrons propagating through the plasma bulk. Figure \ref{fig:DynamicsE55}(f) illustrates that $\updelta$-electrons gain most of their energy in the beginning of the sheath expansion. The newly generated $\updelta$-electrons are accelerated by the sheath electric field and lead to the electron power gain structure inside the sheath ($10\,\rm{ns}\, \leq \, t \, \leq \, 21\,\rm{ns}$).

\begin{figure}[h!]
	\centering
	\includegraphics[width =0.7\textwidth]{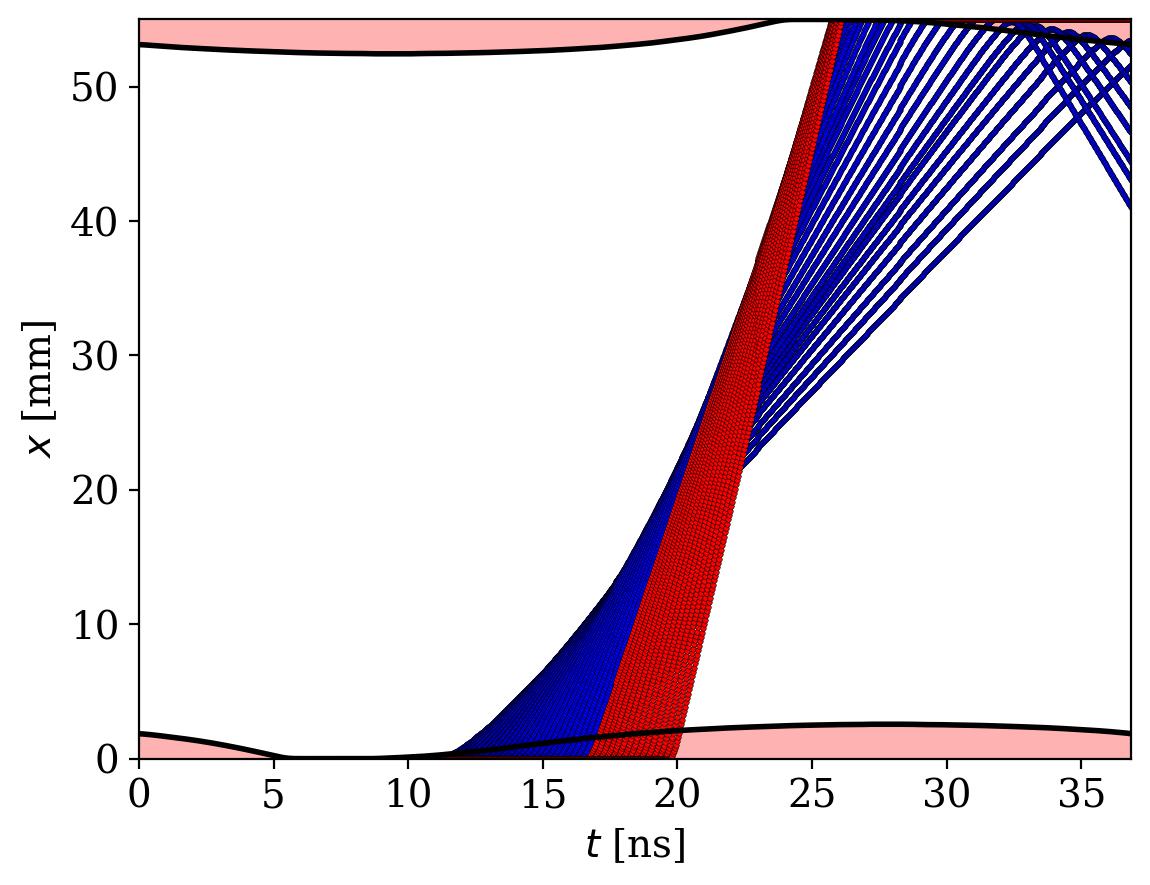}
	\caption{Individual secondary electron trajectories from a PIC/MCC simulation within one RF period. Discharge conditions: $p\,=\,1.0 \  \rm{Pa}\ [\rm{Ar}],$ $V_0\,=\,500 \  \rm{V},$ $f \,=\,27.12\ \rm{MHz},$ $L_{\rm{gap}}\,=\,55 \, \rm{mm}$, SM E. Blue trajectories: SEs which are generated during the beginning of the sheath expansion. Red trajectories: SEs which are generated during the sheath expansion, where a strong sheath electric field is present.}
	\label{fig:Kaustik}	
\end{figure}

The origin of the banana-shaped structure observed for the $\gamma$- and $\delta$-electron dynamics is explained by single particle trajectories. In figure \ref{fig:Kaustik} individual secondary electron trajectories are presented during the sheath expansion, extracted from the PIC/MCC simulation. 
The electrons are generated at slightly different points in time and gain a different amount of energy inside the sheath electric field. Faster secondaries overtake electrons with lower energies (i.e., lower velocity) on their way through the discharge. To visualize this effect, the secondaries are divided into two categories. The blue trajectories indicate electrons born at the beginning of the sheath expansion ($10\,\rm{ns}\, \leq \, t \, \leq \, 17\,\rm{ns}$). Due to the increasingly stronger sheath electric field, the electrons generated later in time overtake those which are born before. Thus, the last-born electron from the blue category hits the opposing sheath first. The red trajectories represent secondaries which are emitted between $17\,\rm{ns}\, \leq \, t \, \leq \, 20\,\rm{ns}$ and have much higher velocities than the blue category. The secondaries do not pass each other, but they are bunched into a narrow beam that impinges first at the opposing sheath. The banana-shaped structure results from the accumulation of electrons (blue and red category) with different velocities. This phenomenon is also known as caustics in optics \cite{causticoptics}. Hence, the banana-shaped structure cannot be described by a macroscopic density but by several individual particles.

\subsection{Electron dynamics at small electrode gap size}
\label{subsection:SmallGap}
\begin{figure}[h!]
	\centering
	\includegraphics[width =\textwidth]{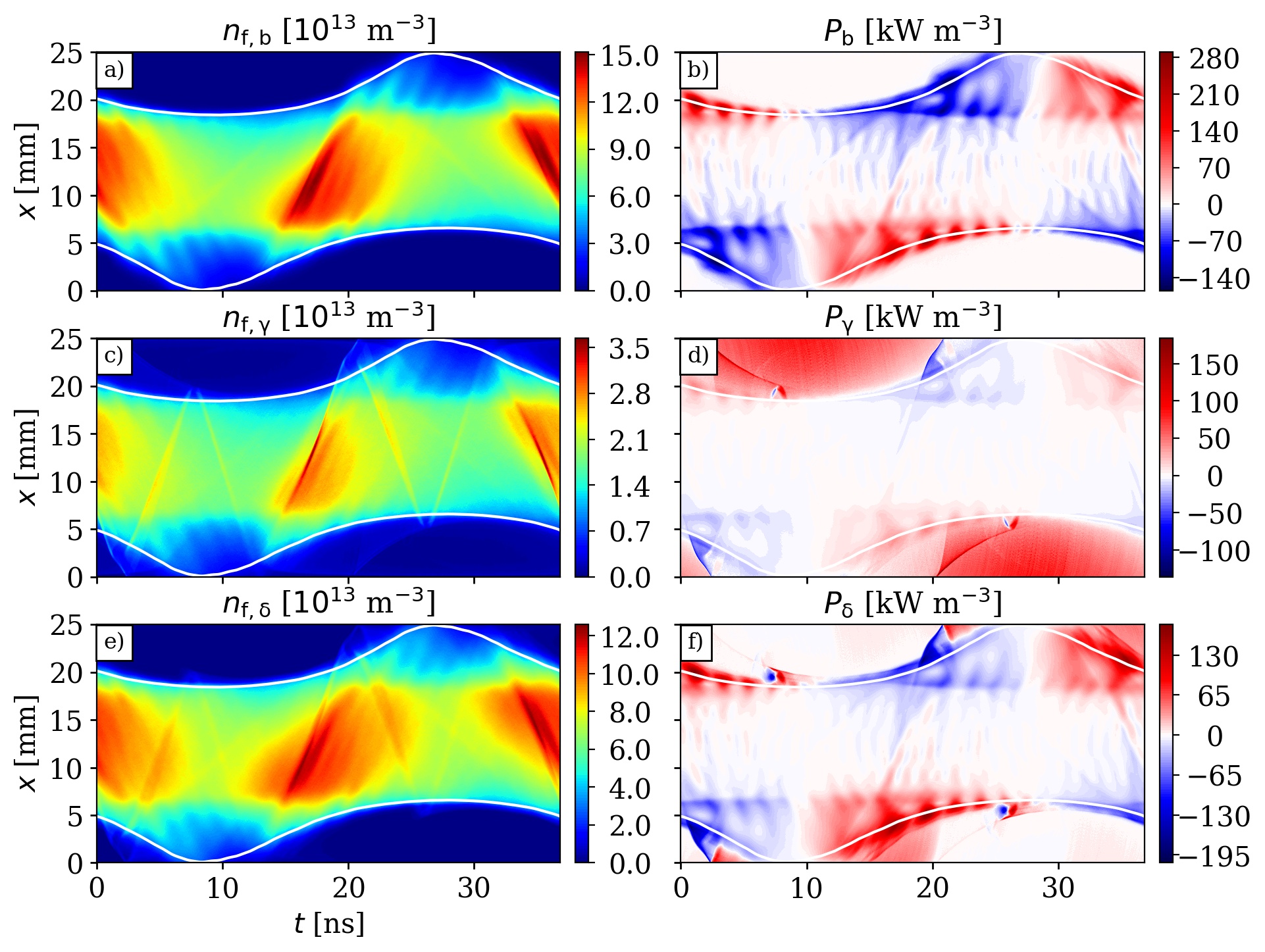}
	\caption{Spatio-temporal distribution of the electron density ($\epsilon\,\geq\,15.8\,\rm{eV}$) and the electron power density of each electron group.
	First column: bulk-electron density (a), $\upgamma$-electron density (c), $\updelta$-electron density (e).
	Second column: bulk-electron power density (b), $\upgamma$-electron power density (d), $\updelta$-electron power density (f). Discharge conditions: $p\,=\,1.0 \  \rm{Pa}\ [\rm{Ar}],$ $V_0\,=\,500 \  \rm{V},$ $f \,=\,27.12\ \rm{MHz},$ $L_{\rm{gap}}\,=\,25\,\rm{mm}$, SM E ($r\,=\,r(\epsilon),\, \gamma\,=\,0.4,\, \delta\,=\,\delta(\epsilon)$).}
	\label{fig:DynamicsE25}	
\end{figure}

Figure \ref{fig:DynamicsE25} presents the same parameters as shown in figure \ref{fig:DynamicsE55}, yet for an electrode gap size of $L_{\rm{gap}}\,=\,25\,\rm{mm}$.
Due to the smaller gap size, most of the fast bulk-electrons (see figure \ref{fig:DynamicsE25} (a)) reach the opposing sheath during the collapsing phase (e.g. top sheath at $20 \leq t \leq 27$ ns).
Consequently, these electrons are decelerated and reflected back into the bulk with less energy.
This deceleration is also shown in figure \ref{fig:DynamicsE25} (b), where the blue structure represents an electron power loss during the sheath collapse.
In addition, the bulk-electrons are able to overcome the sheath potential during the collapsing phase and reach the electrode. 
As a result, electrons are lost at the electrode, generate new $\updelta$-electrons or they are reflected from the electrode. 
Similarly to the result of SM A in figure \ref{fig:ElectronDynamics}(a), this small gap size leads to unfavorable conditions for efficient electron power gain, and thus, leads to a lower electron density \cite{Wilczek_DrivingFrequency,Liu_BRH2011,Liu_BRH2012_2}.
The resulting nonlinear dynamics shown in figure \ref{fig:ElectronDynamics}(b) during sheath expansion are frequently present at low plasma densities and generate multiple fine and distinct structures in the density of fast electrons (generation of multiple electron beams)\cite{Wilczek_Current,Wilczek_resonance,Sharma_Wave,Sharma_Wave2}. However, these nonlinear dynamics do not play a major role in the interplay between bulk, $\upgamma$- and $\updelta$-electrons.

\begin{figure}[h!]
	\centering
	\includegraphics[width =0.7\textwidth]{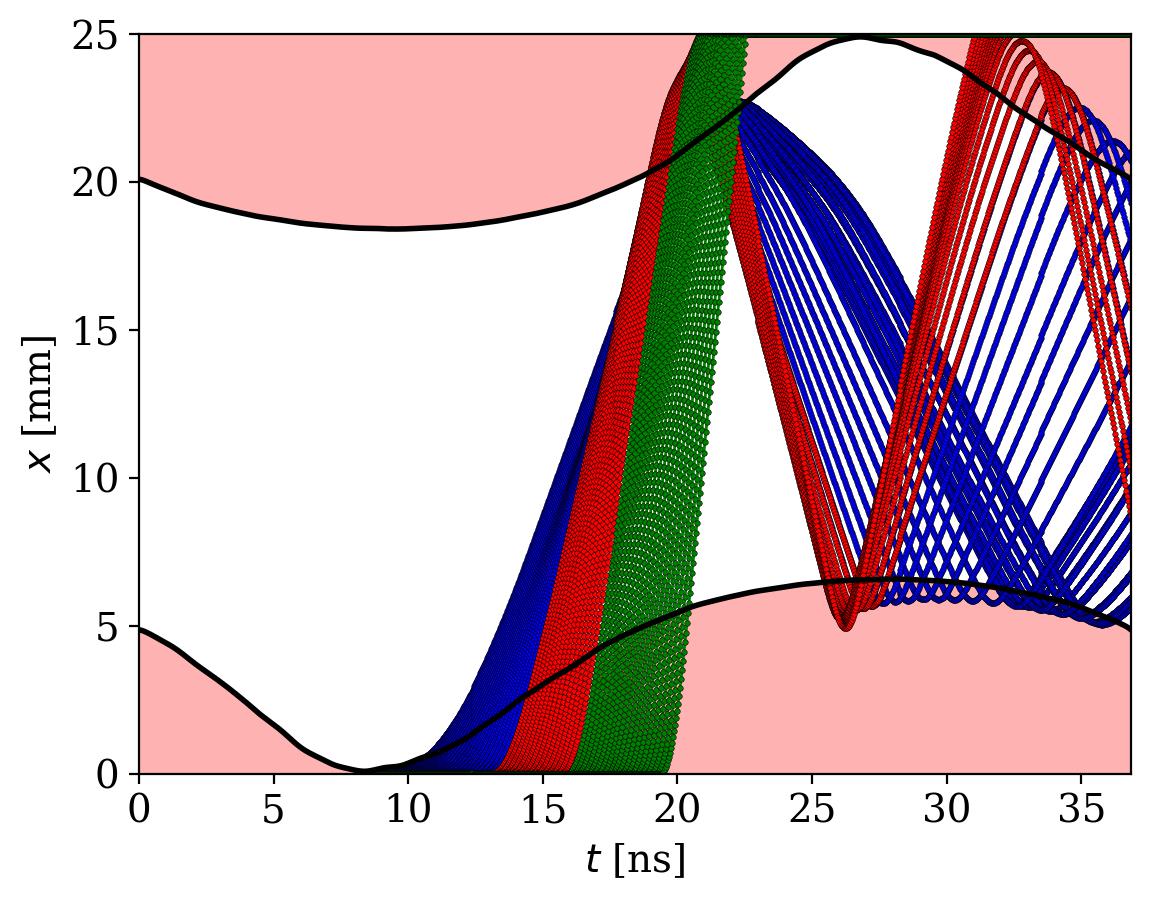}
	\caption{Individual secondary electron trajectories from a PIC/MCC simulation within one RF period. Discharge conditions: $p\,=\,1.0 \  \rm{Pa}\ [\rm{Ar}],$ $V_0\,=\,500 \  \rm{V},$ $f \,=\,27.12\ \rm{MHz},$ $L_{\rm{gap}}\,=\,25 \, \rm{mm}$, SM E. Blue trajectories: SEs which are born during the beginning of the sheath expansion and are reflected within the boundary sheath. Red trajectories: SEs which are born during the sheath expansion, where a strong sheath electric field is present, and are reflected within the boundary sheath. Green trajectories: SEs which are born during the sheath expansion, where a strong sheath electric field is present, and hit the electrode.}
	\label{fig:Kaustik25}	
\end{figure}

The dynamics of the $\upgamma$-electrons in figure \ref{fig:DynamicsE25}(c) can be divided as in figure \ref{fig:DynamicsE55}(c). First, collisions lead to isotropization, and thus, to similar dynamics as in \ref{fig:DynamicsE25}(a) (acceleration during sheath expansion).
Second, the accumulation of $\upgamma$-electrons (see figure \ref{fig:Kaustik25}) leads again to a banana-shaped structure. However, due to the small gap size and the large plasma sheaths, the curvature is minimal and the structure is narrowed.
This accumulated beam of energetic $\upgamma$-electrons penetrates into the upper sheath region at $t\, \approx \, 18.5\, \rm{ns}$. Consequently, these $\upgamma$-electrons are decelerated in the upper sheath electric field, which is shown by the blue structure at $t\, \approx \, 20\, \rm{ns}$ in figure \ref{fig:DynamicsE25}(d). Subsequently, the $\upgamma$-electrons can be reflected inside the sheath or even at the electrode, which leads to a reflected beam moving from the upper sheath to the fully expanded sheath at the bottom electrode (see fig. \ref{fig:DynamicsE25}(c)). Here, the electrons are reflected again, since they cannot overcome the high sheath potential.
However, the essential disparity between the electron dynamics at large and small electrode gap size is that these $\upgamma$-electrons can generate new $\updelta$-electrons if they reach the electrode during the collapsing sheath phase at $t\, \approx \, 18.5\, \rm{ns}$.
As it was shown previously for an electrode gap of 55 mm, the banana-shaped structure penetrated into the sheath during the fully collapsed phase and the generated $\updelta$-electrons gained a few eV. In the $25\,$mm case, however, the newly generated $\updelta$-electrons gain significantly more energy ($\epsilon\,\approx\,100\,\rm{eV}$), which strongly supports the ionization process.    

As already discussed in figure \ref{fig:Matrix} (f), the $\updelta$-electrons have the largest contribution to the ionization rate. This is partly based on the reasoning above that the newly generated $\updelta$-electrons absorb additional energy from the remaining sheath potential resulting in a higher ionization probability. Moreover, the beams of $\updelta$-electrons are effectively confined. Again, the $\updelta$-electrons have the same dynamics as the $\upgamma$-electrons, therefore, they are also reflected two times (see fig. \ref{fig:DynamicsE25}(e)). This leads to an extended path length of the banana-shaped structure compared to the previous case ($L_{\rm{gap}}\,=\,55\,\rm{mm}$), where the banana-shaped structure impinges into the completely collapsed boundary sheath.
This effect increases the ionization probability.
At the temporal phases when the banana-shaped structure of $\updelta$-electrons penetrates into the boundary sheath (top: $t\, \approx \, 20\, \rm{ns}$, bottom: $t\, \approx \, 26\, \rm{ns}$), blue and red structures can be found in figure \ref{fig:DynamicsE25}(f). As already discussed for the $\upgamma$-electrons, the banana-shaped structure is first decelerated (blue) and the remaining electrons are accelerated again (red).
Newly generated  $\updelta$-electrons are accelerated as well ($t\, \approx \, 20\, \rm{ns}$, top electrode).

Figure \ref{fig:Kaustik25} shows individual secondary electron trajectories (representing $\gamma$- and $\delta$-electrons) during the sheath expansion, extracted from the PIC/MCC simulation (similar diagnostic configuration as before, cf. figure \ref{fig:Kaustik}). 
Figure \ref{fig:Kaustik25} displays the electrons divided into three differently colored categories. Similarly, the superposition of electrons with different velocities leads to the banana-shaped structure.
The SEs represented by the blue trajectories are accelerated at the beginning of the sheath expansion ($10\,\rm{ns}\, \leq \, t \, \leq \, 13\,\rm{ns}$) and overtake each other on their way through the plasma bulk. The red category of electrons is bunched into a narrow beam, which hits the opposing sheath before the blue category of electrons.
The interaction with the boundary sheath has a localizing effect on the red electron category bunching the category into a sparse temporal and spatial structure.
The electrons traverse the bulk again and impinge on the fully expanded sheath edge, where they are bunched again and reflected a second time ($4\,\rm{mm}\, \leq \, x \, \leq \, 5\,\rm{mm}$, $26\,\rm{m}\, \leq \, t \, \leq \, 28\,\rm{ns}$).
Since the narrow structure (red category) bunches the electrons, the banana-shaped structure and its two reflected fast electron beams are macroscopically recognizable within the densities (cf. fig. \ref{fig:DynamicsE25}(c) and (e)).
The SEs illustrated by the blue trajectories are also reflected without the localizing effect described above. Their temporal and spatial structure diverges.
The electrons represented by the green trajectories are the fastest, passing the blue electrons without noticeably contributing to the banana-shaped structure. At the temporal position where those electrons arrive at the collapsing boundary sheath, their energy suffices to overcome the sheath potential.
These SEs then have the possibility to be reflected at the electrode, to release an electron-induced secondary electron, or to get lost.

\begin{figure}[t!]
	\centering
	\includegraphics[width =0.7\textwidth]{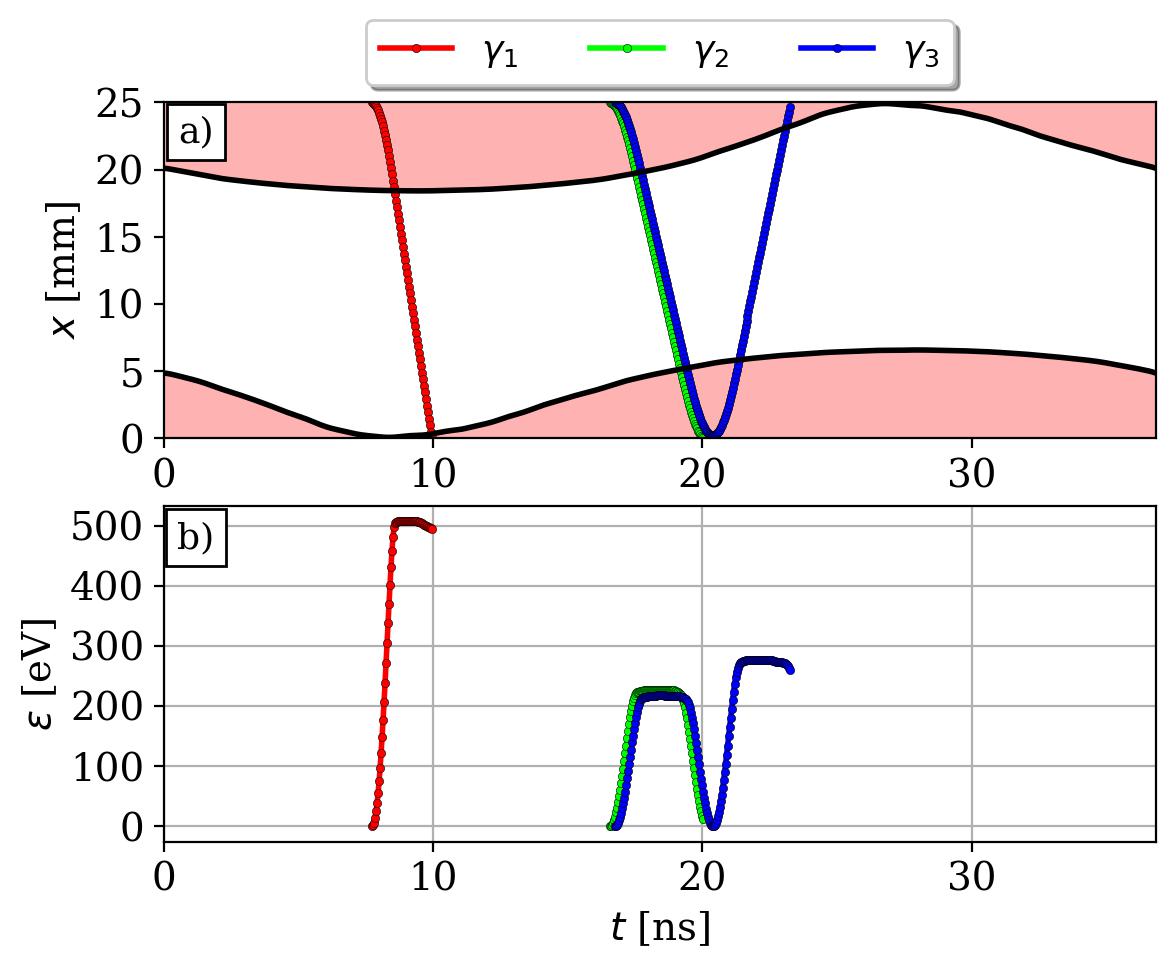}
	\caption{Single $\upgamma$-electron trajectories from a PIC/MCC simulation (a) and the corresponding kinetic energy (b) within one RF period. Discharge conditions: $p\,=\,1.0 \  \rm{Pa}\ [\rm{Ar}],$  $V_0\,=\,500 \  \rm{V},$ $f \,=\,27.12\ \rm{MHz},$ $L_{\rm{gap}}\,=\,25 \, \rm{mm}$, SM E.}
	\label{fig:TestparticleSimulationGammaGamma}	
\end{figure}

According to figure \ref{fig:DynamicsE25}(d), a less intense power gain structure is formed in case of $\gamma$-electrons in the beginning of the sheath expansion ($10\,\rm{ns}\, \leq \, t \, \leq \, 20\,\rm{ns}$) at the powered electrode.
Again, single particle trajectories extracted from the PIC/MCC simulation help the understanding of the origins of this observation.
In figure \ref{fig:TestparticleSimulationGammaGamma} three individual $\upgamma$-electron trajectories (panel (a)) and their corresponding kinetic energy (panel (b)) are shown within one RF period for the same discharge conditions. All the three $\upgamma$-electrons are generated at the grounded electrode at position $\rm{x}\,=\,25\,\rm{mm}$ and are accelerated by the sheath electric field towards the driven electrode. The first $\upgamma$-electron ($\upgamma_1$, red trajectory) gains the maximum sheath voltage, propagates collisionlessly through the bulk, and overcomes the almost minimal sheath potential at the driven electrode around $10\,\rm{ns}$.
Before being absorbed at the electrode, $\upgamma_1$'s energy decreases by the corresponding temporal sheath potential.
The second $\upgamma$-electron ($\upgamma_2$, green trajectory) is generated during a collapsing sheath phase and, therefore, the amount of the kinetic energy gained in the sheath electric field is much less ($230\, \rm{eV}$) than $\upgamma_1$'s.
Simultaneously, the sheath at the driven electrode expands, but $\upgamma_2$ still has enough energy to overcome the sheath potential.
When $\upgamma_2$ reaches the electrode at $20\,\rm{ns}$, it has lost almost all of its kinetic energy. Since all $\upgamma$-electrons $\upgamma_\ast$ created at the grounded electrode during the interval $8\,\rm{ns}\, \leq \, t \, \leq \, 18\,\rm{ns}$ gain a total amount of kinetic energy $\epsilon_{\upgamma_\ast}$ where $\epsilon_{\upgamma_2}\,\leq\,\epsilon_{\upgamma_\ast}\,\leq\, \epsilon_{\upgamma_1}$, all of these electrons carry enough kinetic energy to overcome the sheath potential and reach the powered electrode. 
Thus, the superposition of decelerated $\upgamma$-electrons incoming from the opposing sheath and accelerated $\upgamma$-electrons originating from or reflected at the driven electrode leads to the less intense electron power gain structure at the beginning of the sheath expansion in figure \ref{fig:DynamicsE25}(d).
In contrast,$\upgamma$-electrons generated after $18\,\rm{ns}$ at the opposing sheath can no longer overcome the sheath potential.
This is demonstrated by a third $\upgamma$-electron ($\gamma_3$, blue trajectory) generated shortly after $\upgamma_2$.
Because the sheath at the grounded electrode collapses, $\upgamma_3$ gains less energy ($220\, \rm{eV}$) than $\upgamma_2$ and cannot overcome the sheath potential at the driven electrode.
The previous non-synergistic superposition in the power absorption of $\upgamma$-electrons at the phase of sheath expansion at the powered electrode (seen in figure \ref{fig:DynamicsE25}(d)) does not occur after $20\,\rm{ns}$.
However, $\upgamma$-electrons still penetrate the sheath, as it is shown by figure \ref{fig:TestparticleSimulationGammaGamma}(b): $\upgamma_3$ first loses its kinetic energy when it penetrates the sheath, then it is accelerated backwards instantaneously, gaining energy from the sheath electric field and forming the darker red electron power gain structure observable in figure \ref{fig:DynamicsE25}(d) inside the sheath from $20\,\rm{ns}\, \leq \, t \, \leq \, 36.9\,\rm{ns}$. 
During the sheath expansion at the powered electrode, the penetration depth of incoming $\upgamma$-electrons from the opposing electrode into the sheath decreases. Thus, the dark red structure forming inside the sheath from $20\,\rm{ns}\, \leq \, t \, \leq \, 28\,\rm{ns}$ is quarter-circle-shaped (cf. figure \ref{fig:DynamicsE25}(d)).
The electrons which overcome the sheath potential until $20\,\rm{ns}$ can also generate $\updelta$-electrons.
This is the reason why a positive electron power gain structure at the driven electrode exactly until this point in time can be found in figure \ref{fig:DynamicsE25}(f).

\section{Conclusion}
In this work, the influence of secondary electrons on the overall electron dynamics in low pressure CCRF discharges was studied by means of PIC/MCC simulations.
For this purpose, the electrons were divided into groups of bulk-, $\upgamma$-, and $\updelta$-electrons according to their origin to examine their individual effects on the discharge.
In an electrode gap size variation, it was shown that the gap size controls the impingement phase of high-energy electrons.
Moreover, the details of the electron dynamics are highly influenced by the gap size and the individual electron groups.
Without a remarkable influence of secondary electrons, increasing the gap size causes the electron density to increase until a plateau is reached.
The subsequential increase of the surface coefficients and, thus, the number of secondary electrons causes the plateau in the electron density to disappear and the density to increase (SM D and SM E).
Since the $\updelta$-coefficient may be larger than one, multiple $\updelta$-electrons per incident electrons can be generated.
The resulting population of $\updelta$-electrons is much larger than that of the $\upgamma$-electrons.
In addition, it was found that in the studied regime (i.e., at low pressure, high driving voltage) $\upgamma$- and $\updelta$-electrons largely drive the ionization dynamics.
The high kinetic energy of secondaries (equivalent to the time dependent sheath potential) enables efficient ionization of the background gas.
At smaller electrode gap sizes, more high-energy $\updelta$-electrons are found in the system contributing to the ionization process.
These small gap sizes provide an effective confinement of the secondaries by multiple reflections at the boundary sheath or at the electrodes (see sec. \ref{subsection:SmallGap}).
In contrast, a large electrode gap size makes the $\upgamma$-electrons exhibit an enormous power absorption, leading to a high contribution to the ionization rate.\par
The individual dynamics of high-energy electrons ($\epsilon\,\geq\,15.8\,\rm{eV}$) and their corresponding electron power density were investigated in time and space.
It was shown that an additional beam structure triggered by the SEs is formed.
The full understanding of the additional beams required sophisticated diagnostics on the nanosecond timescale, such as resolving single electron trajectories.
The accumulation of secondary electrons with different velocities was observed, causing SEs passing by each other through the plasma bulk.
The resulting shape of the beam structure was described as banana-like.
In optics, a similar phenomenon is known as caustic.
The electrons inside this banana-shape structure have higher velocities than the beam caused by electrons accelerated at the expanding sheath phase.
The varied electrode gap size serves to control the impingement phase of the banana-shaped structure.
When it hits the beginning of the sheath collapse, the SEs are reflected and then accelerated back into the bulk.
This results in a bouncing effect, where the electrons are reflected several times during one RF period.
This bouncing of SEs is particularly pronounced at small electrode gap sizes explaining the reported excellent electron confinement.
At larger electrode gap sizes the banana-shaped structure hits the fully collapsed sheath phase and triggers the additional generation of $\updelta$-electrons.\par
The analysis of the power gain and loss mechanisms revealed disparities in the dynamics of $\upgamma$- and $\updelta$-electrons.
Ion-induced SEs are continuously released at the electrodes and, thus, gain their energy mainly in the boundary sheath.
Electron induced SEs gain their energy mainly at the beginning of the sheath expansion and secondarily at partial sheath collapse by the residual sheath potential, when the electrode gap is short. The $\updelta$-electrons can solely be generated as long as electrons reach the electrode.
It has been shown that even high energy $\upgamma$-electrons originating from the opposite electrode can overcome the sheath potential and release $\updelta$-electrons just up to a specific point in time.
This interplay was demonstrated to be reflected within the individual energy absorption.
Due to the low pressure and the long electron mean free path of the SEs, the dependency of $\updelta$-electrons on $\upgamma$-electrons is exceptionally high.
Therefore, the electrode gap size, via the impingement phase of high-energy electrons, offers some control over the population of $\updelta$-electrons.\par
We conclude that plasma-surface interactions significantly influence the electron dynamics in capacitively coupled radio frequency discharges.
Neglecting them or using unrealistic coefficients for their treatment results in significant disparities in terms of the simulation results. 
In addition, the electrode gap size serves as a control parameter of the dynamics of both the bulk electrons and SEs.

\section*{Acknowledgments}
\noexpand Financial support provided by the German research foundation in the frame of research projects MU2332 11-1, and 428942393 is gratefully acknowledged.


\printbibliography

\end{document}